\documentclass[journal]{IEEEtran}
\usepackage{cite}
\usepackage{ulem}
\usepackage{bm}
\usepackage{graphicx}
\usepackage{amsmath,bm}
\usepackage{array}
\usepackage{mdwmath}
\usepackage{amssymb}
\usepackage{mdwtab}
\usepackage{stfloats}
\usepackage{amsmath,amsthm}
\usepackage{threeparttable}
\usepackage{color}
\usepackage{ulem}
\usepackage[ruled,linesnumbered]{algorithm2e}
\usepackage{multirow}
\usepackage[caption=false,font=normalsize]{subfig}
\usepackage{fancyhdr}

\makeatletter
\newcolumntype{"}{@{\hskip\tabcolsep\vrule width 3pt\hskip\tabcolsep}}
\makeatother

\begin{document}
	\title{\huge A Novel Dynamic Ray-Tracing Channel Model for 6G LEO Satellite-to-Ground Communication Systems }
	\author{Songjiang Yang, \IEEEmembership{Member,~IEEE}, Cheng-Xiang Wang, \IEEEmembership{Fellow,~IEEE}, Yinghua Wang, \\Jie Huang, \IEEEmembership{Senior Member,~IEEE}, Yuyang Zhou,  Wei Feng, \IEEEmembership{Senior Member,~IEEE}, and\\ el-Hadi M. Aggoune, \IEEEmembership{Life Senior Member,~IEEE}
		\thanks{This work was supported by the Key R\&D Program of China under Grant 2024YFB2907400, the National Natural Science Foundation of China (NSFC) under Grants 62401644 and 62425110, the Major Science and Technology Project	of Jiangsu Province under Grant BG2025039, the Research Fund of National Mobile Communications Research Laboratory, Southeast University, under Grant 2026A05, and the  Promising Researcher Program, University of Tabuk, Saudi Arabia, under Grant PRP-2025-01. This work was presented in part at the IEEE ICCC’24, Hangzhou, China. (\textit{Corresponding Author: Cheng-Xiang Wang})}
		
		\thanks{S. Yang and Y. Wang are with the Purple Mountain Laboratories, Nanjing, 211111, China (email: syang20140111@gmail.com, wangyinghua@pmlabs.com.cn).
			
		C.-X. Wang, J. Huang, and Y. Zhou are with the National Mobile Communications Research Laboratory, School of Information Science and Engineering, Southeast University, Nanjing 211189, China, and also with the Purple Mountain Laboratories, Nanjing 211111, China (email: \{chxwang, j\_huang, yuyzhou\}@seu.edu.cn).
		
		W. Feng is with the State Key Laboratory of Space Network and Communications, Department of Electronic Engineering, Tsinghua University, Beijing 100084, China (e-mail: fengwei@tsinghua.edu.cn).
		
		E. M. Aggoune is with AI and Sensing Technologies Research Center, University of Tabuk, Tabuk 47315/4031, Saudi Arabia (e-mail: haggoune@ut.edu.sa).
			} }
	\markboth{IEEE Transactions on Wireless Communications,~Vol.~XX, No.~XX, Month~2026}
	{Shell \MakeLowercase{\textit{et al.}}: Bare Demo of IEEEtran.cls for Journals}
	\maketitle

\begin{abstract}
	The sixth generation (6G) communication systems envision a space-air-ground integrated network, making accurate satellite-to-ground (S2G) channel modeling crucial for communication systems design. Currently, the majority of S2G channel models rely on stochastic methods, which cannot provide multipath information based on realistic satellite orbits and environments.
	In this paper, we propose a novel low Earth orbit (LEO) S2G channel model based on dynamic ray-tracing to overcome the challenges of tracking rays for long-distance propagation by setting a virtual transmission plane (VTP).
	Moreover, since the line-of-sight propagation mechanism contributes the most power to satellite communications, the VTP is constructed based on the first Fresnel zone to generate parallel rays close to the ground.
	The realistic satellite orbit and environmental information are used in the proposed LEO S2G ray-tracing channel model to compute channel characteristics.
	The results show that the proposed LEO S2G  ray-tracing channel model can accurately generate the channel characteristics based on the satellite orbits. 
	The channel characteristics of LEO S2G communications computed by the proposed ray-tracing method can well match the channel measurements.
\end{abstract}

\begin{IEEEkeywords}
 Channel model, direct-to-cell satellite communications, LEO satellite, multipath fading, ray-tracing
\end{IEEEkeywords}

\IEEEpeerreviewmaketitle

\section{Introduction}
	\normalem
	Driven by the requirements of global coverage in sixth generation (6G) wireless communications, low Earth orbit (LEO) satellite communications have emerged as an essential part of future communication systems. Specifically, the LEO satellite communication systems offer advantages such as low latency, low propagation loss, and flexible networking compared to the high Earth orbit (HEO) and medium Earth orbit (MEO) satellite communication systems~\cite{Heo2023COMST, Lei2025JSAC}. 
	Consequently, LEO direct-to-cell satellite communication systems have garnered much interest from companies SpaceX, OneWeb, and Telesat in recent years~\cite{Curzi2020Aero}.  
	However, to build reliable and efficient 6G LEO satellite-to-ground (S2G) communication systems, it is imperative to possess a comprehensive understanding of channel characteristics for LEO S2G communications~\cite{SYANG2025IoTJ, Lei2025Netw}.
	
	There has been a lot of recent interest in the analysis and design of LEO S2G communication networks \cite{Heo2023COMST, Curzi2020Aero}.
	However, only a handful of  channel models used in these LEO S2G networks can express high accuracy channel characteristics to support the analysis and design of LEO S2G networks.
	To achieve 6G requirements of accuracy in specific environments, the study of LEO S2G ray-tracing channel models is important because they can support detailed multipath information based on the specific environments to design LEO S2G networks.
	
	\subsection{Related Works}	
	Channel measurements are direct and effective ways to know the channel characteristics for specific scenarios \cite{Loo1998Proc}.
	The existing research on S2G channel measurements analyzed the channel characteristics of the satellite channel in different frequency bands, such as L-band (1-2 GHz) \cite{Loo1998Proc, Vogel1995TAP, Cheffena2011TAP}, S-band (2-4 GHz) \cite{Vogel1995TAP, King2007TWC}, Ku-band (12-18 GHz) \cite{Scalise2008TVT, Cid2014TAP, Cid2016TVT}, K-band (18-26.5~GHz) \cite{Goldhirsh1993EL}, and Ka-band (26.5-40 GHz) \cite{Papafragkakis2019AWPL}.
	Moreover, the channel measurements for S2G communications with different elevation angles were conducted in the European area \cite{King2007TWC}.
	The ground user in LEO satellite communications suffered from strong variations of the received power due to the shadowing and multipath fading because of direct path and reflection paths from objects in the surroundings~\cite{Lutz1991TVT}.
	When the S2G link between the satellite and the ground user experienced no clear line-of-sight (LoS), the shadowing of the signal had the main effect on LEO satellite communications.
	The channel characteristics of satellite communications were analyzed for channel measurements in urban, suburban, and maritime scenarios, such as root mean square (RMS) delay spread, K factor, coherence bandwidth, average fading duration~\cite{Cheffena2011TAP, Vogel1995TAP, Scalise2008TVT, Cid2014TAP, Cid2016TVT, Wang2026TWC}.
	Despite extensive channel measurements across diverse frequency bands and scenarios for S2G communications, not all scenarios are easily measured and analyzed because of the high cost.
	
	According to S2G channel measurements, LEO S2G communications have some unique characteristics, such as the Faraday rotation, ionospheric scintillation, and atmospheric influence, in complex environments \cite{CXWang2023COMST}.  
	The Faraday rotation predominantly affects the power of non-circularly polarized and low-frequency electromagnetic waves. 
	Conversely, circularly polarized electromagnetic waves remain impervious to the power ramifications of Faraday rotation \cite{Vucetic1992JSAC}.
	The ionospheric scintillation introduces variations in amplitude and phase, with relatively minor impacts on lower frequency bands~\cite{3gpp2017study}. 
	The atmospheric influence, characterized by the effects of cloud, rain, and fog, was formulated as the different cloud, rain, and fog attenuation in many different situations~\cite{Su2024IoTJ}. 
	Existing literature expounds on these phenomena, offering well-established methodologies to model the extra attenuation of the LEO S2G communications.
		
	Currently, the channel models are generally divided into stochastic and deterministic channel models. 
	Geometry-based stochastic models (GBSMs) can be adapted to various scenarios and frequency bands through parameter adjustments, where the channel characteristics can be computed by considering the geometry relationship in the propagation environments~\cite{CXWang2022TVT, CXwang2023Tcom, CXWang2025SCIS}. 
	In \cite{Chen2024TWC}, the impact of spacecraft on the satellite channel model was considered with LoS and multipath from solar-panel.
	Quasi-deterministic Radio Channel Generator (QuaDRiGa) was a well-known tool to simulate land mobile satellite communications with distance, azimuth spread of arrival/departure, and elevation spread of arrival/departure based on GBSM. 
	However, QuaDRiGa did not fully consider the Faraday rotation and estimate the attenuation in real-time for S2G communications \cite{Burkhardt2014EuCap}.
	A LEO S2G large-scale fading channel model based on LoS probability and GBSM model was proposed by considering the ground reflection and the LoS path \cite{Su2024IoTJ}.
	A LEO satellite-assisted deep space channel model was first proposed in \cite{Gao2025TAE}, which can effectively characterize the fading condition.
	In \cite{Abdi2003TWC} and \cite{Alsseur2008TWC}, the satellite channel models were proposed to express the unique channel characteristics based on some fading model or multi-state Markov chain.
	Although GBSM can model the accurate statistics channel characteristics of the LEO S2G communications with general environment parameters, the accuracy of channel state information for some specific propagation environments may not achieve the level of 6G requirements.
	
	Ray-tracing, a type of deterministic channel model, offers higher accuracy based on geometric optics (GO) and the uniform theory of diffraction (UTD) \cite{SYANG2024TVT, He2019TUT}.
	Channel models in 6G necessitate precise channel modeling for deterministic scenarios due to the requirements of high transmission rate and low latency for 6G networks.
	In~\cite{Dottling2001APM}, the two-dimensional~(2D) and three-dimensional (3D) wideband land mobile satellite ray-tracing channel models were considered, where the received power for different cases was shown. However, how the ray-tracing keeps the accuracy after the long-distance transmission was not mentioned. In~\cite{Oestges2001TVT}, a ray-tracing tool for satellite-to-urban communications was used to analyze the transmission strategy. The losses due to the diffraction and reflection were considered in the ray-tracing, and the delay power spectral density (PSD) and path loss prediction were discussed. In~\cite{Yan2019ACCESS}, a satellite-to-vehicle link was analyzed using ray-tracing to evaluate the Rician K-factor, delay spread, and angular spreads in both urban and highway scenarios. The satellite-to-helicopter channel model based on the 3D ray-tracing was proposed and compared with the measurements, where the power level fading and Doppler PSD due to the helicopter rotor were analyzed \cite{Cid2014TAP}. In the measurement, the transmitter was set at a tower 15 m high to act as the satellite transmitter, and the helicopter was placed on the ground, where similar satellite-to-helicopter measurement scenarios were rebuilt in the ray-tracing simulation. In these works, the LEO satellite was set as the static point in the air with a specific elevation angle and only considered the ground receiver mobility. We note that how the realistic satellite trajectories were combined in the ray-tracing has not been sufficiently studied.
		﻿
		
	Based on the static setting of the satellite in existing S2G ray-tracing channel models, a plane near the ground was proposed to emit the ray for considering the reflection mechanism, but how to set the plane was not mentioned~\cite{Nicolas2012EuCap}. Furthermore, since the high directional antenna was equipped on the satellite and fewer scatterers existed on the satellite side, the whole propagation path was divided into satellite and ground parts in a LEO S2G channel model for urban scenarios~\cite{Zhang2024EuCap, Ning2024ICCC}. The rays were emitted from a plane near the ground to compute multipath components accurately based on the urban environment. However, the relationship between the satellite trajectories and the near-ground area was not defined in these methods. We note that the ray propagation for the long-distance and how the near ground plane setting in the ray-tracing has not been sufficiently studied.
	
	\subsection{Contributions}
	In this paper, we propose a LEO S2G channel model based on dynamic ray-tracing, incorporating the satellite trajectories and the parallel rays arriving at the ground user. 
	More specifically, the channel characteristics between the LEO satellite and the ground user are derived by using the coordinate transformation and the proposed equivalent Earth model. 
	Moreover, a virtual transmission plane (VTP) is constructed to emit the ray to maintain the parallel ray assumption for the LEO S2G communications and to address the long-distance propagation challenge inherent in the ray-tracing. 
	The main contributions and novelties of this paper are summarized as follows.
	\begin{itemize}
	\item 
	A coordinate transformation for LEO satellite orbits is used to define the unified coordinate system between the LEO satellite and the ground user. 
	An equivalent Earth model is proposed to generate the elevation angle between the LEO satellite and ground link, and the effective propagation distance for the troposphere and ionosphere. 
	The ionospheric scintillation and atmospheric effects are considered in the large-scale fading model based on the relative position information.
	Moreover, the small-scale fadings, e.g., the Doppler PSD, are derived and computed based on the proposed equivalent Earth model.
	\item 
	A VTP is constructed as a new ray-emitting plane in the LEO S2G communications to transfer the original point source rays from the satellite into parallel rays near the ground. 
	Moreover, the height and radius of the VTP are determined based on the first Fresnel zone and the environmental geometry. 
	The VTP can realize the parallel ray propagation in the near-ground, a unique characteristic of LEO S2G communications, and overcome the long-distance propagation problem between the satellite and the ground user, which helps the receiver capture more significant multipath components.
	\item 
	A pre-processing method tailored for the LEO S2G communications is proposed to combine with the dynamic ray-tracing concept and reduce computational complexity by considering the time-varying elevation angle of the LEO satellite. 
	The ranges of rays received are classified by the elevation angle of the LEO satellite with different reflection and diffraction propagation mechanisms, which can adapt to the propagation environments. 
	Accordingly, the angle ranges of rays received can be used to compute the ray creation and suppression times of the traditional dynamic ray-tracing.  
\end{itemize}  

	The rest of this paper is organized as follows. 
	In Section~II, the LEO S2G channel model for ray-tracing with the coordinate transformation, VTP, and pre-processing method is presented. 
	In Section III, how to compute the channel characteristics of LEO S2G communications is described.
	In Section IV, the results of the proposed LEO S2G ray-tracing channel model are shown. 
	Finally, in Section V, our main conclusions are drawn.

\section{ A Dynamic Ray-tracing for S2G Communications}
	LEO S2G communications differ significantly from terrestrial communications in terms of ray-tracing, particularly in defining the propagation environment and modeling ray emission.
	The LEO satellite is navigating in orbit around the Earth, and the length of available transmission time between the LEO satellite and the ground user is dependent on the relative position of the LEO satellite and the ground user.	
	Moreover, the details of the ray-tracing for LEO S2G communications and the pre-processing method to reduce the computational complexity are presented.
	
\begin{table*}
	\centering
	\caption{List OF Abbreviations.}
	\begin{tabular}{|c|c|c|c|}
		\hline
		Abbreviations & Full text& Abbreviations & Full text \\
		\hline
		2D & two-dimensional & 3D & three-dimensional \\
		\hline
		3GPP & Third-generation partnership project & 6G & sixth generation \\
		\hline
		6GPCM & 6G pervasive channel model & AAoA & Azimuth angle of arrival \\
		\hline
		CPU & Center processing  unit &EAoA & Elevation angle of arrival \\
		\hline
		ECEF & Earth-centered Earth-fixed &ENU & East-North-Up  \\
		\hline
		  GBSM & Geometry-based stochastic model&GO & Geometric optics \\
		\hline
		 GPS & Global positioning system& GPU & Graph processing unit \\
		\hline
		HEO & High Earth orbit & LEO & Low Earth orbit \\
		\hline
		LLA & Latitude, Longitude and Altitude & LoS & Line-of-sight \\
		\hline
		MEO & Medium Earth orbit & PSD & Power spectral density \\
		\hline
		QuaDRiGa & Quasi-deterministic Radio Channel Generator & RMS & Root mean square \\
		\hline
		SBR & Shooting and bouncing ray & STK & System Tool Kit \\
		\hline
		UTD & Uniform theory of diffraction & VTP & Virtual transmission plane \\
		\hline
	\end{tabular}
\end{table*}

	\subsection{LEO Satellite Orbit and Coordinate Transformation}
\begin{figure}[t]
		\centering
		\includegraphics[width=0.7\linewidth]{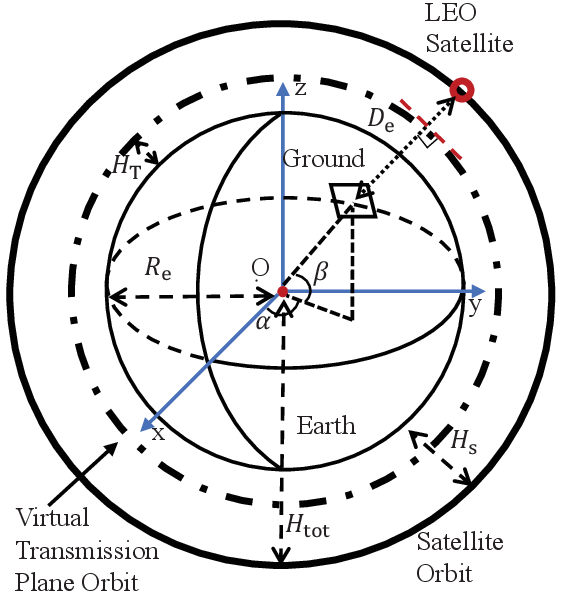}
		\caption{An equivalent Earth model for LEO S2G communications.}
		\label{fig:LEOearth}
	\end{figure}

	There are many data formats to express the satellite orbit with different coordinate systems, such as two-line element data, J2000 coordinate system, and Latitude, Longitude and Altitude (LLA) coordinate system \cite{Shirazi2021AESM}.
	Moreover, the LEO satellite orbit can also be generated with the simplified general perturbations algorithm with orbit data in a different coordinate system \cite{Ning2024ICCC}, where the information of satellite orbit can be found from the open database of the launched satellite\cite{Celes}.
	To maintain generality, the LLA coordinate system is chosen to be used in the proposed method because it is widely used in the global positioning system (GPS).
	However, the LLA is a geodetic coordinate, which makes it difficult for the ray-tracing to get the relative position between the LEO satellite and the ground user.
	Hence, the LLA coordinate needs to be converted to the Earth-centered Earth-fixed (ECEF) coordinate system, which is a simple Cartesian coordinate system at the center of the Earth. 
	The ECEF coordinate of the Earth is used to build the equivalent Earth model, which is shown in Fig. \ref{fig:LEOearth}.
	The parameters of the equivalent Earth model are described in Table~\ref{tab:para}.
		
	To convert LLA coordinate system into ECEF coordinate system, the following conversion equations \cite{Hofmann-Wellenhof1997} can be used with the LLA coordinate location information to get the location of the satellite ($x^{'}_\mathrm{s}, y^{'}_\mathrm{s}, z^{'}_\mathrm{s} $) in ECEF coordinates,
	\begin{equation}\label{eq:conversionX}
		x^{'}_\mathrm{s}=(N(\Lambda)+ H_\mathrm{s})\cos(\Lambda)\cos(\psi)
	\end{equation}
	\begin{equation}\label{eq:conversionY}
		y^{'}_\mathrm{s}=(N(\Lambda)+ H_\mathrm{s})\cos(\Lambda)\sin(\psi)
	\end{equation}
	\begin{equation}\label{eq:conversionZ}
		z^{'}_\mathrm{s}=((1-\Xi)^2N(\Lambda)+ H_\mathrm{s})\sin(\Lambda)
	\end{equation}
	\begin{equation}\label{eq:conversionN}
		N(\Lambda) = \frac{R_\mathrm{e}}{\sqrt{1-\Xi^2(2-\Xi)\sin^2(\Lambda)}} 
	\end{equation} 
	where $ \Lambda $ is the value of latitude, $ \psi $ is the value of longitude, $ H_\mathrm{s} $ is the height of satellite trajectory, $\Xi$ is the flattening of Earth, and $ R_\mathrm{e} $ is the effective radius of Earth. 
	The ECEF coordinate system is a standard coordinate system used to transform between other coordinate systems in the satellite geometry computation.
		
	\begin{table}[t]
		\caption{Parameters of equivalent Earth model.}
		\centering
		\begin{tabular}{|c|l|}
			\hline
			Parameters & Definitions  \\
			\hline
			\multirow{2}{*}{$H_\mathrm{tot}$} & The height between the center of Earth and LEO\\ 
			& satellite orbit  \\
			\hline
			$ H_\mathrm{T} $& The height of the VTP\\
			\hline
			$ H_\mathrm{s} $ & The height of the LEO satellite orbit \\
			\hline
			\multirow{2}{*}{$ D_\mathrm{e} $}& The effective distance between LEO satellite and\\
			& ground user \\
			\hline
			$ R_\mathrm{e} $ & The effective radius of Earth  \\
			\hline
			\multirow{2}{*}{$ \alpha$}& The azimuth angle of the center of Earth and\\
			& ground user \\
			\hline
			\multirow{2}{*}{$ \beta $}& The elevation angle of the center of Earth and \\
			& ground user  \\
			\hline
		\end{tabular}
		\label{tab:para}
	\end{table}
	For efficient use in ray-tracing, the ECEF coordinate system is converted into a local coordinate system to represent the position of ground user and the propagation environments. 
	The East–North–Up (ENU) coordinate system is commonly adopted for this purpose, as it defines the local tangent plane and facilitates the calculation of angles and propagation geometry.
	To convert the ECEF coordinate system to the ENU coordinate system, the location of ground user ($ x_\mathrm{g}, y_\mathrm{g}, z_\mathrm{g} $) can be set as a local reference point and the location of LEO satellite ($x^{'}_\mathrm{s}, y^{'}_\mathrm{s}, z^{'}_\mathrm{s} $) is the target point in the ECEF coordinate~\cite{Hofmann-Wellenhof1997}.
	The transformation coordinate is shown as
	\begin{equation}
		\begin{bmatrix}
			x_\mathrm{s} \\
			y_\mathrm{s} \\
			z_\mathrm{s}
		\end{bmatrix}
		\!=\! 
		\begin{bmatrix}
			\!-\!\sin \Lambda_\mathrm{g} & \cos\Lambda_\mathrm{g} & 0 \\
			\!-\!\cos \Lambda_\mathrm{g}\sin\psi_\mathrm{g} & \!-\!\sin\Lambda_\mathrm{g}\sin\psi_\mathrm{g} & \cos\Lambda_\mathrm{g} \\
			\cos \Lambda_\mathrm{g}\cos\psi_\mathrm{g} & \sin\Lambda_\mathrm{g}\cos\psi_\mathrm{g} & \sin \psi_\mathrm{g}
		\end{bmatrix}
		\begin{bmatrix}
		x^{'}_\mathrm{s}\!-\! x_\mathrm{g}	\\
		y^{'}_\mathrm{s}\!-\! y_\mathrm{g}\\
		z^{'}_\mathrm{s}\!-\! z_\mathrm{g}	
		\end{bmatrix}
	\end{equation}
	where $\Lambda_\mathrm{g} $  and $\psi_\mathrm{g} $ are the latitude and longitude of the ground user, respectively.
	Accordingly, the position of the ground user and LEO satellite in ENU coordinate are expressed as $(x_\mathrm{g}, y_\mathrm{g}, z_\mathrm{g}) $ and $(x_\mathrm{s}, y_\mathrm{s}, z_\mathrm{s}) $, respectively.
	The effective distance between LEO satellite and ground user is computed by
	\begin{equation}\label{equ:effDistance}
		D_\mathrm{e}=\sqrt{(x_\mathrm{s}-x_\mathrm{g})^2+(y_\mathrm{s}-y_\mathrm{g})^2+(z_\mathrm{s}-z_\mathrm{g})^2}.
	\end{equation} 
	The azimuth angle ($\theta$) and elevation angle ($\gamma $) in ENU coordinate system between the LEO satellite and ground user are computed by
	\begin{equation}\label{eq:theta}
			\theta=\tan^{-1}\left(\frac{y_\mathrm{s}-y_\mathrm{g}}{x_\mathrm{s}-x_\mathrm{g}}\right)
	\end{equation}
	\begin{equation}\label{eq:gamma}
		\gamma=\sin^{-1}\left(\frac{z_\mathrm{s}-z_\mathrm{g}}{D_\mathrm{e}}\right).
	\end{equation}
	Finally, the LEO satellite and ground user are in the same coordinate system, where the relative distance and angle between the LEO satellite orbit and ground user trajectories can be computed by the ray-tracing.
		
\subsection{Ray-tracing with Virtual Transmission Plane} 
	Due to the long-distance propagation between the LEO satellite and ground user, an accurate and effective LEO S2G channel model becomes challenging by using traditional ray-tracing methods, such as shooting and bouncing ray (SBR) and image methods.  
	For the traditional SBR method, the rays emitted by the LEO satellite transmitter are based on an icosahedron, and these rays have a large angular dispersion at the receiver area near the ground for the LEO S2G communications because of the long-distance propagation.
	Accordingly, even if the angular dispersion is very small at the transmitter side, it will become larger at the receiver side, and the ray cannot interact with the scatterers on the ground near the user.
	Moreover, since the image method is based on the geometry concept, where the scale of the environment and objects needs to be similar, it is difficult to use in the LEO S2G communications.   
	Hence, some modifications of the existing ray-tracing are needed to overcome the long-distance propagation in the LEO S2G communications.
	
\begin{figure}[t]
		\centering
		\includegraphics[width=0.8\linewidth]{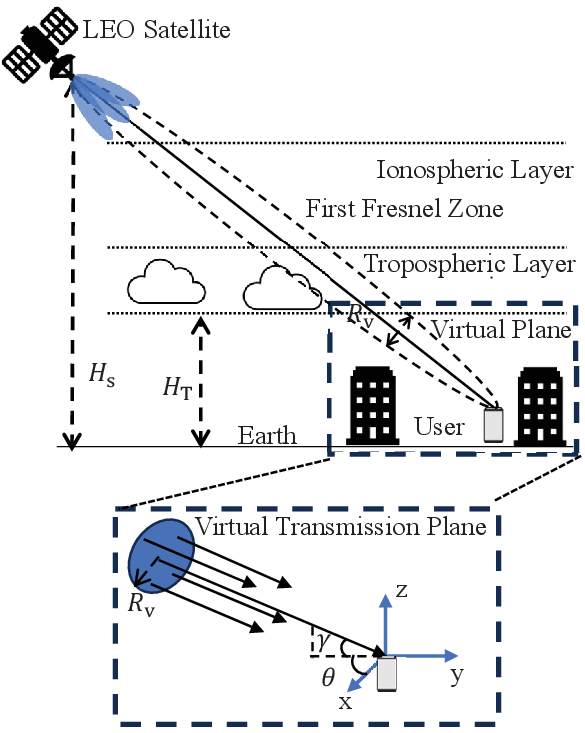}
		\caption{LEO S2G links with the VTP.}
		\label{fig:LEOfresnel}
	\end{figure}

	To address these issues, a VTP is designed near the ground side to facilitate the process of tracing the ray in the proposed LEO S2G ray-tracing channel model, which is depicted in Fig.~\ref{fig:LEOfresnel}.
	In the real propagation process, the electromagnetic wave transmitted from the LEO satellite is undergoing directional antenna and beamforming with directivity, resulting in multiple narrow and directional beams \cite{Heo2023COMST}. 
	Moreover, the LEO S2G link mainly relies on the line-of-sight (LoS) propagation, and the multipath effects happen near the ground \cite{Dottling2001APM}.
	Accordingly, the VTP is set as a near-ground virtual plane for the ray emission of the LEO S2G ray-tracing channel model to emulate the propagation characteristics of the beam between the LEO satellite and the ground user.
	In addition, the VTP ensures the parallel wave assumption near the surface of the Earth, reducing deviations of the emitting ray from the point source used in the traditional ray-tracing.
	
	Since the majority of the propagating energy resides within the first Fresnel zone for a LoS link \cite{Dottling2001APM} and the LEO S2G link mainly relies on the LoS propagation, the first Fresnel zone is a vital reference for the VTP setting.
	Accordingly, the radius of the first Fresnel zone can be set as the radius of the VTP.
	The radius of the first Fresnel zone is given as
	\begin{equation}\label{equ:firstFresnel}
		R_\mathrm{v}=\sqrt{\frac{\lambda (H_\mathrm{s}-H_\mathrm{T})H_\mathrm{T}}{H_\mathrm{s}}}
	\end{equation}
	where $ \lambda $ is the wavelength of the carrier frequency, $ H_\mathrm{s} $ is the height of the LEO satellite orbit, and $ H_\mathrm{T} $ is the height of the VTP. 
	According to \eqref{equ:firstFresnel}, the minimum radius of the VTP depends on the height of the VTP set in the LEO S2G communications. 
	The maximum radius of the VTP is constrained by the beam footprint of the S2G link, but the large radius of the VTP will increase the computational cost for the proposed method.
	
	In the LEO S2G link, the various factors of obstruction, such as encompassing urban structures, satellite elevation angles, and atmospheric conditions within the Fresnel zone, can also impact the propagating energy \cite{Liolis2010TVT}. 
	Since the propagation mechanism for LEO S2G needs to maintain the LoS as the majority propagation mechanism, the height of the VTP needs to be much higher than buildings to keep the high LoS probability~\cite{SYANG2022TVT, Pang2023IoTJ}. 
	Moreover, atmospheric conditions in the troposphere include clouds, rain, and fog, where the large-scale fading includes not only rain attenuation but also partial obstruction of the Fresnel zone by fog.
	It is noteworthy that the obstruction in the troposphere primarily occurs at the wave propagation energy level rather than the ray-tracing propagation level. 
	To simplify the model and eliminate the influence of atmospheric rays, the height of the VTP can be set below the height of the troposphere.
	Hence, the height range of the VTP is between the height of buildings and the height of the troposphere.
	The height of the VTP not only facilitates the consideration of ray-tracing when buildings obstruct rays emitted from the VTP, but also results in easy computation of multipath for the near-ground side.
	
	Variations in LEO satellite elevation angles are also a crucial factor influencing the Fresnel zone.
	During low elevation angles, the components of direct and reflected paths exhibit diminished impact, necessitating consideration of diffraction effects. 
	The Fresnel zone is a crucial consideration for diffraction propagation, guiding pertinent diffraction propagation scenarios.
	To express the effects of the LEO satellite elevation angles, the VTP is set perpendicular to the azimuth angle and elevation angle between the LEO satellite and the ground user, which is shown in Fig. \ref{fig:LEOearth}.
	Accordingly, the VTP is moving based on the trajectories of the LEO satellite, and the height of the VTP is set under the Tropospheric layer.
		
	For the ray emission process, the ray sources are uniformly distributed on the VTP based on the size of the VTP.
	The minimum space of the ray is consistent with the radius of the reception sphere at the receiver.
	Each ray source emits one ray with the same propagation direction to keep the parallel rays from the VTP.
	Moreover, the rays emitted from the VTP are perpendicular to the plane, so the angle of the rays is the same as the angular information of LEO S2G communications.
	
	In LEO S2G communications, the ground receiver is often affected by clutter effects, e.g., buildings and vegetation \cite{Yu2025TAP}. In the proposed model, these clutter effects are captured through the integration of the VTP and specific environments in ray-tracing. The multipath fading, shadowing, and blockage effects are computed based on the interaction between parallel rays and the specific geometry of the ground environment. Consequently, the proposed method can model the clutter effects in LEO S2G communications.
	 	
	\subsection{Pre-processing Method for Dynamic Ray-tracing}
	The LEO satellite trajectory is an important part of the LEO S2G ray-tracing channel model, but it will cause a high computational burden for the ray-tracing because of the many snapshots from the satellite trajectory. Currently, dynamic ray-tracing is used to reduce complexity for vehicular networks with high mobility by mathematical extrapolation. Although the original dynamic ray-tracing is not directly applicable to LEO S2G communications, the concept of extrapolation can be adapted to account for the dynamic trajectory of the LEO satellite. Specifically, the existing pre-processing for dynamic ray-tracing is used for the linear trajectories to find the extrapolating period by computing the ray creation and suppression points~\cite{SYANG2024TVT}. However, since the satellite is navigating in circular orbits, the pre-processing method needs to be adapted to the circular trajectories by considering the angular change for the satellite.
	
\begin{figure}[t]
		\centering
		\subfloat[Original method]{\includegraphics[width=0.49\linewidth]{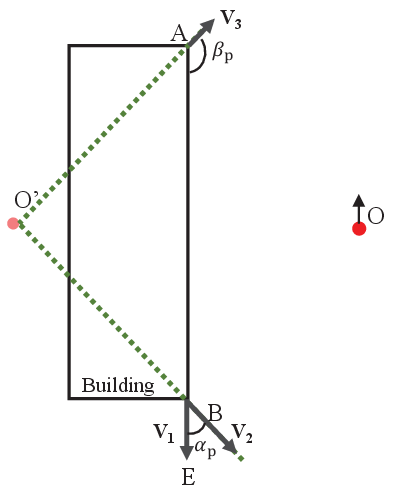}}
		\hfil
		\subfloat[Proposed method]{\includegraphics[width=0.49\linewidth]{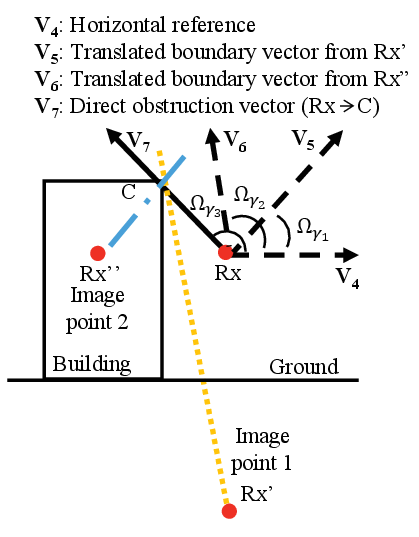}}
		\hfil
		\caption{A schematic diagram of determining angular bouncing ranges for LEO S2G communications, (a) original method \cite{SYANG2024TVT}, (b) proposed method.}
		\label{fig:angledivision}
	\end{figure}

	In the existing pre-processing method, the angular bouncing range can be referenced as the image points or received points with the building edge, as shown in Fig. \ref{fig:angledivision}(a). Moreover, the angular bouncing range is determined by the geometry between the transmitter (or its image point) and the building edge, with the interaction point serving as the reference for calculating the angular range. For example, the angular bouncing range is defined by $\alpha_\mathrm{p}$ and $\beta_\mathrm{p}$ to compute the ray creation and suppression point in Fig. \ref{fig:angledivision}(a), where the first reflection angle range of point O is defined. 
	
	To adapt the LEO S2G dynamic ray-tracing channel model, the angular bouncing ranges are divided by the elevation angle among receiver point (Rx), image receiver points (Rx' and Rx''), and obstruction points (C) on various edges of buildings, as shown in Fig. \ref{fig:angledivision}(b). 
	It is noted that the angular bouncing range needs to be referenced as the receiver point in the proposed LEO S2G ray-tracing channel model because of the parallel wave from the VTP, which is different from the existing method. Therefore, the angular bouncing range for the proposed S2G ray-tracing channel model can be transferred to the receiver point from the image points and compared with the elevation angle $ \gamma $. For the classification of the angular bouncing range, the original line from the image receiver points to point C is translated to the receiver points, where the vectors $\mathbf{V}_5$ and $\mathbf{V}_6$ in Fig. \ref{fig:angledivision}(b) are used to compute the angular bouncing range. The lines Rx''C and Rx'C are parallel to $\mathbf{V}_5$ and $\mathbf{V}_6$, respectively. The value of the angular bouncing range at the receiver point with different conditions can be computed following the basic geometry theorem with reference vector $\mathbf{V}_4$. In Fig. \ref{fig:angledivision}(b), the angular bouncing range is defined by $ \Omega_{\gamma_l} $, where $l$ denotes the elevation angle of satellite in the different angular range, $l \in \{1, 2 , \cdots\}$.
	Let compute the angular bouncing ranges~$ \Omega_{\gamma_1} $ between $\mathbf{V}_4$ and $\mathbf{V}_5$ as an example,
	\begin{equation} \label{eq:angular}
		\Omega_{\gamma1}=\cos^{-1} \left[ \frac{\mathbf{V_4} \cdot \mathbf{V_5}}{|\mathbf{V_4}||\mathbf{V_5}|} \right].
	\end{equation}
	By comparing the angular bouncing range and elevation angle, the considered propagation mechanisms of the LEO S2G ray-tracing channel model in different ranges are summarized as follows.
	\begin{itemize}
		\item When  $0^{\circ} \leq \gamma < \Omega_{\gamma_1}$, the range encompasses direct, ground-reflected, building-reflected, and diffracted paths.
		\item When  $ \Omega_{\gamma_1} \leq \gamma < \Omega_{\gamma_2} $, direct, ground-reflected and diffracted paths are observed.
		\item When  $\Omega_{\gamma_2} \leq \gamma < \Omega_{\gamma_3}$, the region consists of direct and diffracted rays.
		\item When $\Omega_{\gamma_3} \leq \gamma \leq 180^{\circ}$, the region is associated with diffracted paths.
	\end{itemize}
	Since the LEO S2G communications with long-distance propagation, the paths combining direct path, diffraction up to first order, and reflections up to the third order contain most of the transmission energy \cite{Nikolaidis2018TAP}.
		
\begin{figure}[t]
		\centering
		\includegraphics[width=0.6\linewidth]{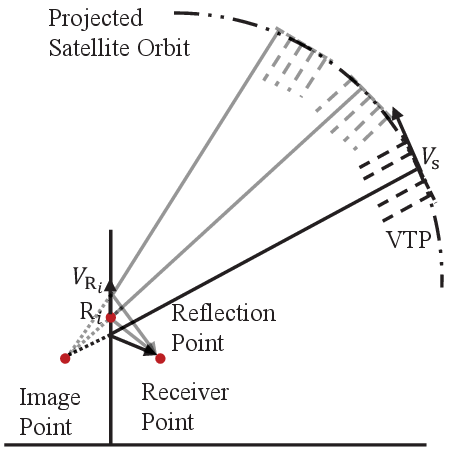}
		\caption{A schematic diagram of the dynamic ray-tracing with time-varying angular for LEO S2G communications.}
		\label{fig:Dynamic}
	\end{figure}

	The elevation angle of satellite in the different angular range~$l$ can be used to compute the ray creation and suppression time as follows
	\begin{equation}\label{eq:coherence}
		T_{\mathrm{c},l}=\frac{\Omega_{\gamma l}}{V_{\mathrm{s},l}}
	\end{equation}
	where $	T_{\mathrm{c},l} $ is the ray creation and suppression time for range~$ l $,  and $V_{\mathrm{s},l}$ is the angular velocity of the satellite.
	Accordingly, the changing information of the ray in this stable period can be computed based on the geometry using an analytic extrapolation method and not trace the ray again because the period is stable for the ray creation or suppression \cite{Bilibashi2023TAP}.
	In the ray creation and suppression time, the interaction points can be computed by the vertices of the chain as shown in Fig. \ref{fig:Dynamic}, where the position and velocity of reflection points are denoted as $\mathrm{R}_i $ and $ V_{\mathrm{R}_i}$, respectively. 
	In the original dynamic ray-tracing process, the transmitter is a point source to generate the rays, and the velocity is the linear representative. 
	However, in the proposed method, the parallel rays are generated by the VTP, which is different from the original.
	The parallel rays from the VTP will change the angle based on the orbit, and the radius of the VTP is fixed for the simulation, where these settings may cause the loss of the ray from the VTP moving.
	To avoid this situation, the reflection ray with the VTP moving will continue to exist before the angular bouncing range is changed, even if the reflection ray is out of the radius of the VTP.
	The angular velocity of the satellite could be converted to the linear velocity by considering the total length from the center of the Earth to the orbit in the proposed method, where the concept of the analytic extrapolation is consistent with the original dynamic ray-tracing in \cite{Bilibashi2023TAP}.
	The linear velocity of the VTP moving in dynamic ray-tracing is shown as 
	\begin{equation}
		V_{\mathrm{s,linear}} = V_\mathrm{s}(R_\mathrm{e}+H_\mathrm{T}).
	\end{equation}
	After converting the linear velocity of the satellite, the extrapolation computation of the ray information follows the dynamic ray-tracing process.
	
\begin{figure}[t]
		\centering
		\includegraphics[width=0.7\linewidth]{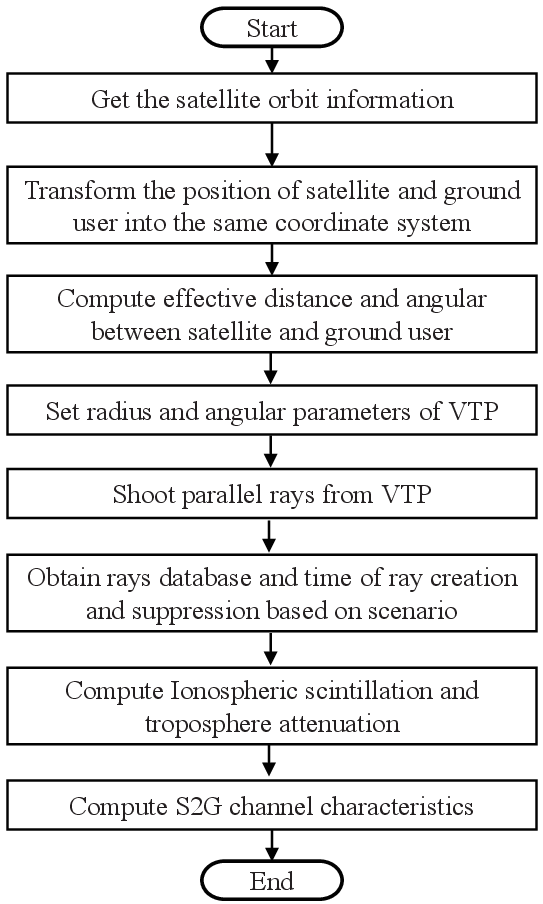}
		\caption{A flowchart of dynamic ray-tracing for LEO S2G communications.}
		\label{fig:LEOflowchart}
	\end{figure}

	The flowchart of the whole LEO S2G ray-tracing channel modeling is shown in Fig.~\ref{fig:LEOflowchart}.
	The main steps include getting the LEO satellite orbit information, coordinate transformation for the LEO satellite and ground user, setting the VTP, shooting parallel rays from VTP with the dynamic ray-tracing concept, obtaining the ray database, and computing the LEO S2G channel characteristics.
	
	For scenarios involving large LEO satellite constellations, the proposed channel model can be applied with different satellite orbits in the constellation, and the same ground environment with the ground node. Hence, only the satellite orbits are replaced with those of a different satellite in the proposed channel model to allow for significant data reuse. Furthermore, to overcome the computational burden for a large LEO satellite constellation, parallel acceleration methods based on central processing unit (CPU) and graphics processing unit (GPU) need to be used \cite{SYANG2024TVT, He2019TUT}.
	
\section{Channel Characteristics of LEO S2G Communications}
	In this section, the channel impulse response of the proposed LEO S2G ray-tracing channel model is derived.
	Moreover, the channel characteristics of LEO S2G communications and their impact on the proposed channel model are analyzed.
	
	\subsection{Channel Model}
	Using the ray-tracing, the complex channel impulse response at time $t$ and carrier frequency $ f_\mathrm{c} $ of LEO S2G channel model can be represented by \cite{You2020JSAC}
	\begin{equation}\label{eq:CIR}
		h(t,f_\mathrm{c})=\sum_{i=1}^I g_i e^{j2 \pi\left(f_\mathrm{c}t+f_{\mathrm{D},i}t-f_\mathrm{c}\tau_i\right)} \delta\left(t-\tau_i\right)
	\end{equation}
	where $I$ is the number of paths between the satellite to ground user, $g_i$ is the path amplitude of $i$th path  ($ i \in \{1, 2,...,I\} $), $ f_{\mathrm{D},i} $ is the Doppler shift of $i$th path, $ \tau_{i}$ is the delay of $i$th path, and $ \delta(\cdot) $ is the Dirac delta function.
	This channel impulse response can be adapted to different propagation scenarios, where the parameters in~\eqref{eq:CIR} are dependent on the specific scenarios with the different propagation mechanisms, materials, and simulation settings in the ray-tracing.
	The main parameters in~\eqref{eq:CIR} affect the large-scale and small-scale fading channel characteristics in the path loss, delay, and Doppler. 
	The details of some channel characteristics between the LEO satellite and ground users, such as the path loss, RMS delay spread, and Doppler PSD, are described as follows.
	
	\subsection{Path Loss}
	In LEO S2G communications, the total path loss between the LEO satellite and the ground user considers the propagation attenuation from different propagation mechanisms and extra attenuation based on the propagation environment.
	The propagation attenuation is computed following the electromagnetic computation method for the propagation mechanism with different scenario settings, such as LoS, reflection, and diffraction.    
	Moreover, the extra attenuation, including the atmospheric and ionospheric scintillation, is considered in the received power for the ground user.
	
	Accordingly, the total path loss between the  LEO satellite and ground user is computed by
	\begin{equation}
		g_\mathrm{s} = g_{\mathrm{e}} + g_{\mathrm{Ion}} + g_{\mathrm{Atm}}
	\end{equation}
	where $ g_{\mathrm{e}} $ is the propagation attenuation of the LEO S2G communications, $g_{\mathrm{Ion}}$ is the ionospheric scintillation attenuation, and $ g_{\mathrm{Atm}} $ is the atmospheric attenuation.
	The propagation attenuation including LoS, reflection and diffraction, is computed by
	\begin{equation}\label{equ:total}
		g_{\mathrm{e}}=\frac{|\sum_{i=1}^{l}E_i|^2}{2Z_0}A_\mathrm{r}
	\end{equation}
	where $ E_i $ is electric fields for $i$th path, $ Z_0 $ is the characteristic impedance of the air, $ A_\mathrm{r}$ is the aperture of the antenna.
	The electric fields for $i$th path  $ E_i $ is computed by the ray-tracing following the GO and UTD, where the reflection and diffraction coefficients are dependent on the realistic scenarios~\cite{He2019TUT}.
	Hence, the path amplitude of $ i$th path $ g_i $ can also be computed following \eqref{equ:total} by only counting one ray. 
	
	The ionospheric scintillation attenuation $g_{\mathrm{Ion}}$ is mainly influenced by the total attenuation at the carrier frequency below 3~GHz, and it can be determined as \cite{3gpp2017study}	
	\begin{equation}
		g_{\mathrm{Ion}} = 27.5 \times \upsilon^{1.26}
	\end{equation}
	where $\upsilon$ is a amplitude scintillation index factor.
	The amplitude scintillation index is computed by intensity fluctuations of the signal and can be scaled by the frequency.
	At frequencies above 3~GHz, the atmospheric effect has significant influence on the total attenuation of the LEO S2G communications.
	Especially, when carrier frequency is above 10~GHz, the rain attenuation becomes heavy for LEO S2G communications.
	The atmospheric attenuation is computed by \cite{3gpp2017study}
	\begin{equation}
		g_{\mathrm{Atm}} = \kappa R^\alpha 
	\end{equation}
	where $ R $ is rain rate in mm/h, $\kappa$ and $\alpha$ are functions of operating frequency, respectively.
	Furthermore, $\kappa$ and $\alpha$ are computed by
	\begin{equation} 
		\label{equ: kpapa}
		\log _{10} \kappa =\sum_{j=1}^{4} a_{j} \exp \left[-\left(\frac{\log _{10} f-b_{j}}{c_{j}}\right)^{2}\right]+m_{k} \log _{10} f+c_{k}
	\end{equation}
	and
	\begin{equation}
		\label{equ: alphapara} 
		\alpha=\sum_{j=1}^{5} a_{j} \exp \left[-\left(\frac{\log _{10} f-b_{j}}{c_{j}}\right)^{2}\right]+m_{\alpha} \log _{10} f+c_{\alpha}
	\end{equation}
	where $ a_{j} $, $ b_{j} $, $ c_{j} $, $ m_{\alpha} $,  $m_{k}$, $ c_{k} $, and $ c_{\alpha} $ are different sets of coefficient to compute the atmospheric attenuation for different electromagnetic wave polarization, which is shown in \cite{3gpp2017study}. 
	For simplicity, the ionospheric scintillation and atmospheric attenuation are unchanged in one loop of ray-tracing, but they can be updated for different loops of ray-tracing. 
		

	\subsection{RMS Delay Spread}
	The delay of each path $\tau_{i}$ is computed by the total length of each path $D_{i}$, which consists of the distance at the near ground with the different propagation mechanisms $D_{\mathrm{G},i}  $ and the distance between the LEO satellite and VTP.
	The distance at the near ground with the different propagation mechanisms $D_{\mathrm{G},i}$ is obtained from the ray-tracing.
	Accordingly, the delay of each path $\tau_{i}$ can be computed as
	\begin{equation}
		\tau_{i}=\frac{H_{\mathrm{S}}-H_{\mathrm{T}}+D_{\mathrm{G},i}}{\mathrm{c}}.
	\end{equation}	
	Hence, the RMS delay spread $\tau_{\mathrm{RMS}}$ is defind as
		\begin{equation}\label{eq:RMSdelay}
		\tau_{\mathrm{RMS}}=\sqrt{\frac{\sum_{i=1}^I{g_i\tau_{i}^2} }{\sum_{i=1}^Ig_i} -\left( \frac{\sum_{i=1}^I{g_i\tau_{i}} }{\sum_{i=1}^Ig_i}\right)^2}. 
	\end{equation}

	\subsection{Doppler PSD}
	In LEO satellite communications, the Doppler shift for propagation path $ i $ of the ground user can be impacted by the motions of the LEO satellite and the ground user.
	The Doppler shift induced by the LEO satellite remains effectively identical across all propagation paths because of the long-distance propagation between the LEO satellite and ground users.
	The Doppler shift arising from ground user mobility in LEO satellite communications is governed by the local scattering geometry surrounding the ground user \cite{You2020JSAC}.
	However, only the Doppler shift caused by the motions of the LEO satellite is considered because the velocity of the LEO satellite is much faster than that of the ground user.
	
\begin{figure}[t]
		\centering
		\includegraphics[width=0.7\linewidth]{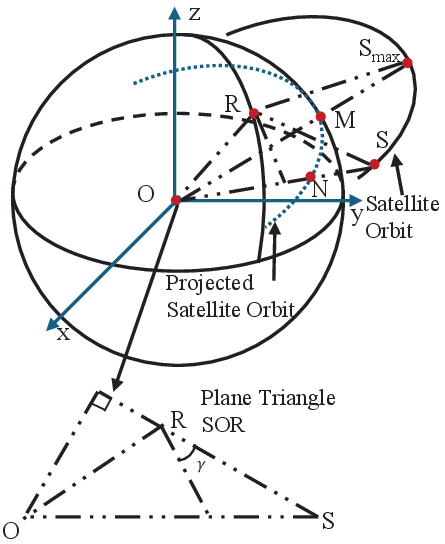}
		\caption{An equivalent Earth model for S2G communications to compute the Doppler information.}
		\label{fig:Doppler}
	\end{figure}

	Fig.~\ref{fig:Doppler} shows the geometric diagram of a LEO satellite and ground user, where point R denotes the ground user, point S denotes the position of satellite first observed for the ground user, point $\mathrm{S}_{\max}$ denotes the position of the satellite when it reaches the maximum elevation angle that ground user can observe, point M denotes the point on projected satellite orbit for the maximum elevation angle, and point N the point on projected satellite orbit for the first observation.
	According to the geometry in Fig. \ref{fig:Doppler}, the elevation angle can be calculated by the basic theorem of triangles and the law of cosines as
	\begin{equation}
		\gamma=\pi / 2 -\angle \mathrm{ROS} -\angle \mathrm{RSO}.
		\label{eq:angel}
	\end{equation}
	For a given orbit of satellite and ground user, the maximum elevation angle and the minimum elevation angle are obtained by calculating the elevation angle at each moment.
	Based on the geometry in Fig. \ref{fig:Doppler}, the Doppler shift in the ECEF coordinate system can be obtained by the maximum elevation angle, which is proposed by \cite{Ali1998Tcom},
	\begin{equation}
		f_{\mathrm{D}}(t)=-\frac{f_{\mathrm{c}}r_{\mathrm{E}}r\sin (\psi_0-\psi_{\max}))\cos \omega_{\max}) V}{\mathrm{c}\sqrt{r_{E}^{2}+r^{2}-2r_{E}r\cos (\psi_0 -\psi_{\max})\cos \omega_{\max}} }    
		\label{eq}
	\end{equation}
	where $r_{\mathrm{E}}$ is the radius of the earth, \textit{r} is the distance from the satellite to the center of the earth, $\psi_0 -\psi_{\max}$ is the angular distance between points M and N, $V$ is the angular velocity of satellite, and $\omega_{\max}$ can be calculated by
	\begin{equation}
		\omega_{\max}=\cos^{-1} \left(\frac{r_\mathrm{E}}{r}\cos \gamma_{\max}\right)-\gamma _{max} 
		\label{eq}
	\end{equation}
	where $\gamma_{\max}$ is the maximum elevation angle.
	Based on calculations above, $t_{\mathrm{du}}$ is related to the maximum elevation angle and the minimum elevation angle
	\begin{equation}
		t_{\mathrm{du}}\approx \frac{2}{V-V_{\mathrm{E}}\cos \eta}\cos^{-1} \left(\frac{\cos \omega_{0}}{\cos \omega_{\max} } \right)      
		\label{eq}
	\end{equation}
	where  $V_{\mathrm{E}}$ is the angular speed of the earth's rotation, $ \eta $ is the orbital inclination, and $\omega_{0}$ is the angular for point S.

\section{Results and Analysis}
	The proposed LEO S2G ray-tracing channel model is validated by two case studies for large-scale and small-scale fading.
	In Case~1, a realistic roadside scenario is considered~\cite{Cheffena2011TAP}, where the results from traditional and proposed methods are compared with measurement data to validate the accuracy of the proposed method.
	In Case~2, the RMS delay spread, path loss, and Doppler shifts are obtained with the simulations in an urban environment.
	In this case, the VTP is following the realistic satellite orbit to generate some channel characteristics.
	Moreover, the comparison of computational complexity for the proposed method with and without the pre-processing method is shown by using the Case 2 environment. 

	\subsection{Case 1: Roadside Scenarios}	
\begin{figure}[t]
		\centering
		\includegraphics[width=0.8\linewidth]{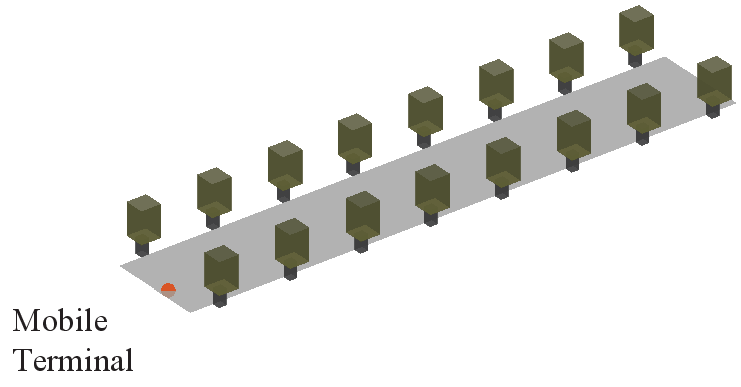}
		\caption{Simulation scenario for Case 1.}
		\label{fig:TreePL}
	\end{figure}

	In 2006, the French Space Agency  conducted S-band propagation measurements in Gaillac, France, to characterize the narrowband channel in controlled environments like tree-lined roads. 
	A mobile terminal moved along a 70~m straight path positioned 6~m from rows of trees flanking both sides of the road. 
	The trees, spaced 9~m apart on each side, formed seven pairs that periodically shadowed the mobile terminal during its trajectory. 
	The transmitter used a 2~GHz signal and was oriented perpendicular to the mobile’s path while maintaining the 30$^{\circ}$ elevation angle. 
	These measurements, detailed in \cite{Cheffena2011TAP}, aimed to validate propagation models in this environment, focusing on shadowing effects caused by the tree pairs. 
	The roadside scenario for simulation is built, and the mobile terminal receiver is set in the middle of the road, which is shown in Fig. \ref{fig:TreePL}.
	
\begin{figure}[t]
		\centering
		\includegraphics[width=\linewidth]{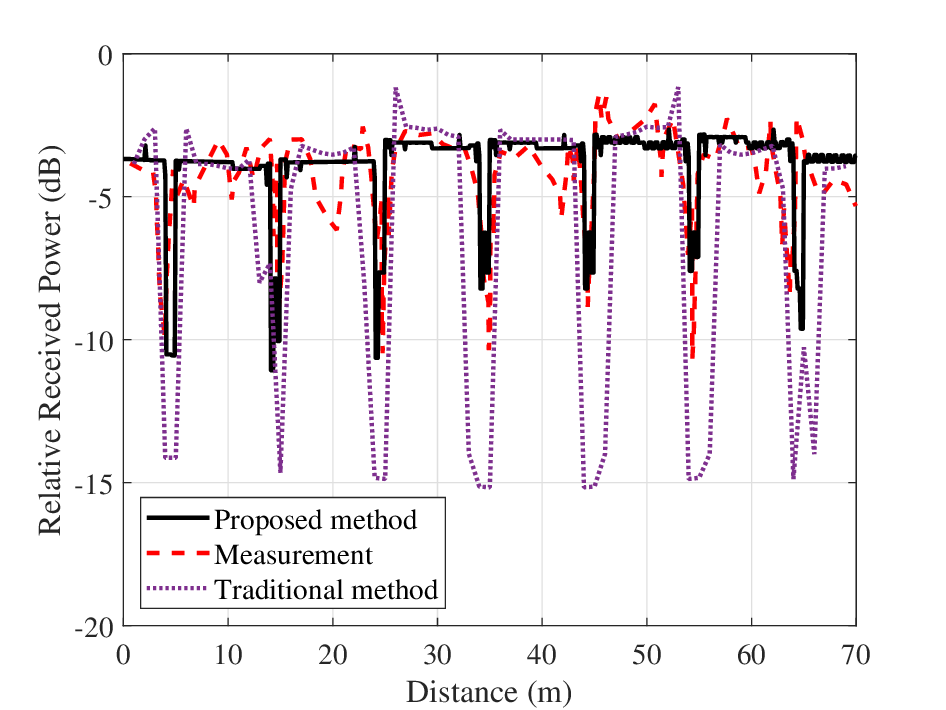}
		\caption{A comparison of path losses among the measurement, proposed method, and traditional method \cite{Oestges2001TVT}.}
		\label{fig:SatPL}
	\end{figure}

	According to \cite{Cheffena2011TAP}, the VTP is set at a 30$^{\circ}$ elevation angle, and it is moving along with the mobile terminal to transmit the parallel ray.
	The traditional method following~\cite{Yan2019ACCESS} is compared, where the transmitter is set at the middle of the scenario with the same position of the VTP.
	Moreover, the other simulation parameters are set following the measurement settings.
	The relative received powers for the measurement, the proposed method, and the traditional method in Case~1 are shown in Fig. \ref{fig:SatPL}, where the measurement results are normalized with respect to the free-space received power. 
	In general, the value of relative received power from the proposed method is in good agreement with the measurement result.
	The mean absolute error and RMS error of relative received power between the measurement and proposed method are 1.19 dB and 1.76 dB, respectively.
	Since the tree model is simplified in the simulation, some missing rays from the realistic tree cause the difference in received power.
	The relative received power of the traditional method is lower than that of the proposed method because the number of rays received at the receiver in the traditional method is less than that of the proposed method. 
	Moreover, the distance blocked by the tree is similar in the proposed method and measurements.
	For the traditional method, the relative received power at the blocked position is symmetrical, which is not similar to the measurement.
	
	\subsection{Case2: Manhattan City Simulation}
\begin{figure}[t]
		\centering
		\includegraphics[width=0.7\linewidth]{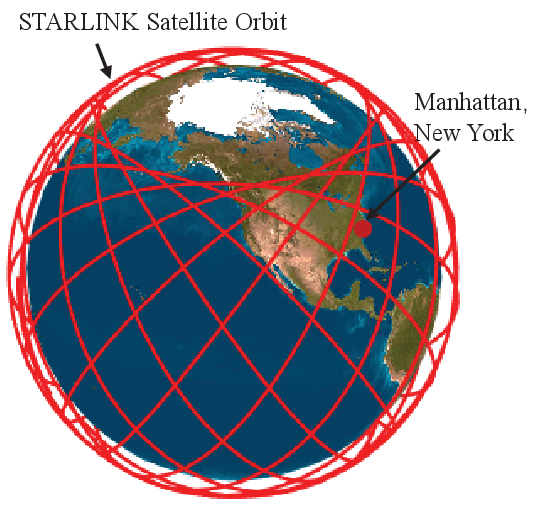}
		\caption{STARLINK LEO satellite orbit in the simulation.}
		\label{fig:Manhattan_sat}
	\end{figure}

	In this case, the orbits of STARLINK-4105 and STARLINK-5240 are simulated in Manhattan city to evaluate the proposed S2G ray-tracing channel model performance.
	The orbital parameters of STARLINK-4105 and STARLINK-5240 are obtained from the open-access database \cite{Celes}
	Based on these parameters, the visible time of both satellites in Manhattan are recorded and the complete orbit of STARLINK-4105 is shown in Fig. \ref{fig:Manhattan_sat}.
	The Manhattan city scenario is rebuilt following the guidelines in the CloudRT database \cite{Yan2019ACCESS, Zhou2025TAP}.
	The detailed simulation parameters can be found in Table \ref{tab:mytable}.
	In all simulations, the dynamic ray-tracing with the VTP and the proposed pre-processing method are used. 
	
\begin{figure}[t]
	\centering
	\subfloat[Traditional method]{\includegraphics[width=0.4\textwidth]{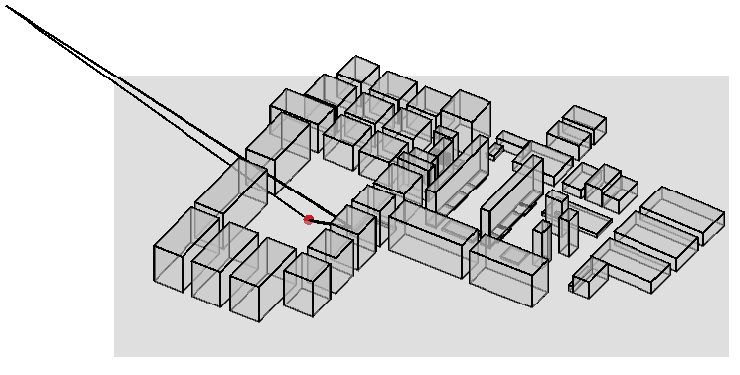}}
	\hfil
	\subfloat[Proposed method]{\includegraphics[width=0.4\textwidth]{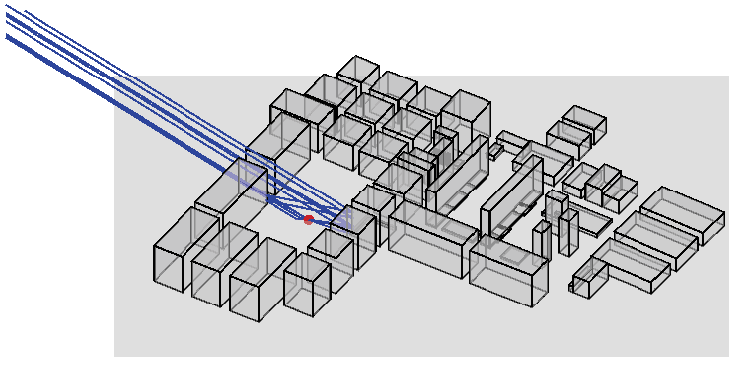}}
	\hfil
	\caption{Paths illustration of the proposed and traditional methods in Manhattan city, (a)  traditional method \cite{Oestges2001TVT}, (b) proposed method.}
	\label{fig:MahaSce}
	\end{figure}

	The ray propagation of the traditional and proposed methods in the Manhattan city simulation scenario using the STARLINK-4105 orbit is shown in Fig.~\ref{fig:MahaSce}.
 	In this example, reflection paths from the satellite to the ground receiver are traced up to the third order.
	The transmitter is set following the STARLINK-4105 orbit with the same height of the VTP in the traditional method~\cite{Yan2019ACCESS}.
	The receiver point location is (-55, 265) and the height is 1.5 m, where the receiver point is in the middle of the open square.
	According to Fig. \ref{fig:MahaSce}, the parallel ray is realized by the VTP, where the ray with the similar propagation paths in the traditional method can also be captured by the proposed method.
	The number of rays for the traditional method is less than that of the proposed method, where more details of the LEO S2G channel can be captured.
	Therefore, the parallel ray setting is important to capture the electromagnetic wave propagation for the LEO S2G ray-tracing channel model.

	\begin{table}[t]
	\caption{Simulation parameters.}
	\centering
	\begin{tabular}{|c|c|}
	\hline
	Parameters & Values  \\
	\hline
	Carrier frequency (\textit{$f_\mathrm{c}$}) & 2, 14, 28 GHz \\
	\hline
	Transmission power (\textit{$P_\mathrm{t}$}) & 30 dBm \\
	\hline
	Minimum satellite altitude (\textit{H}) & 542 km \\
	\hline
	Satellite angular velocity (\textit{$w_\mathrm{s}$}) & 0.066 rad/min  \\
	\hline
	Height of VTP (\textit{$H_\mathrm{T}$}) & 7 km  \\
	\hline
	\end{tabular}
	\label{tab:mytable}
	\end{table}
	\begin{figure}[t]
	\centering
	\includegraphics[width=0.9\linewidth]{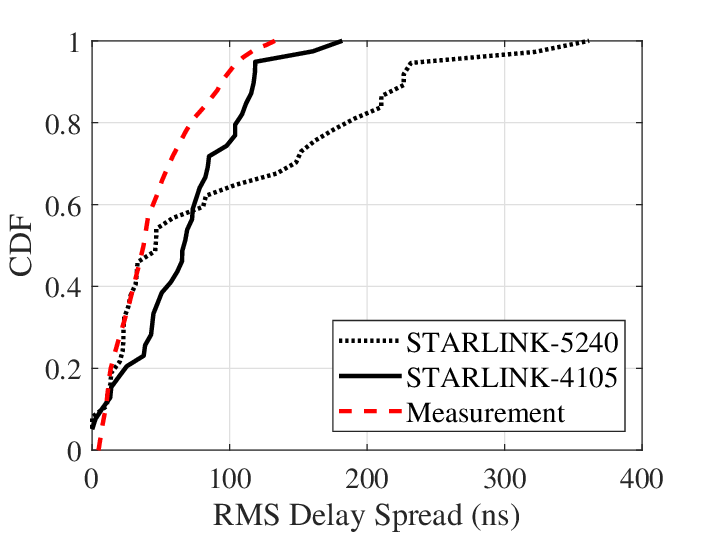}
	\caption{A comparison of RMS delay spreads among the STARLINK-4105, STARLINK-5240, and measurement \cite{Cid2016TVT} in Manhattan city at 14 GHz.}
	\label{fig:RMSDS}
\end{figure}

	\begin{figure}[t]
	\centering
	\includegraphics[width=0.9\linewidth]{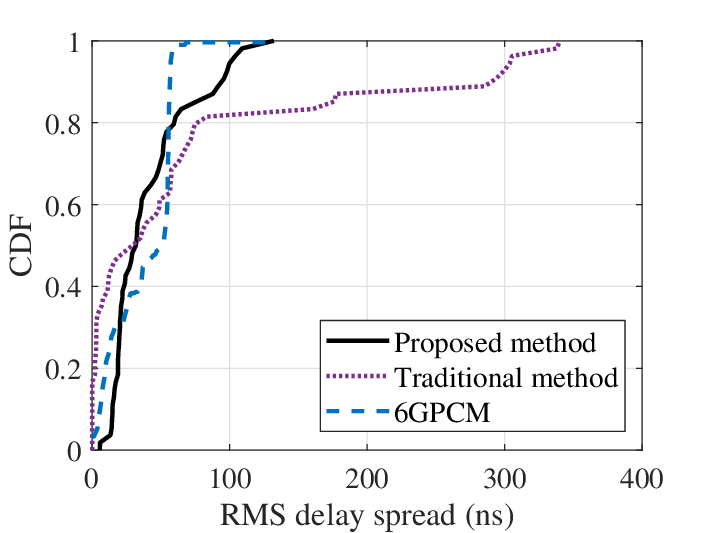}
	\caption{ A comparison of RMS delay spreads among the proposed method, traditional method \cite{Oestges2001TVT}, and 6GPCM \cite{CXWang2022TVT} in Manhattan city at 14 GHz.}
	\label{fig:RMSDSComp}
\end{figure}

	The RMS delay spread of the LEO S2G communications in an urban scenario, obtained from measurement~\cite{Cid2016TVT}, is compared with the results of the proposed method for different satellite orbits, as shown in Fig.~\ref{fig:RMSDS}. 
	The measurements were conducted by using a sweep time delay cross-correlation sounder operating at 14 GHz in an urban area.
	The sounder was placed on the roof of a building with the elevation angles between 10$^{\circ}$ and 14$^{\circ}$, and the distance between the transmitter and receiver was 180-240 m.
	Since channel characteristics statistics are consistent with similar scenarios~\cite{Seyedi2013TWC}, the simulation results for the STARLINK-4105 and STARLINK-5240 in Manhattan city can be compared with the urban measurement results.
	In practical scenarios, the elevation angle range of the LEO S2G communications is not from 0 to 90$^{\circ}$ because the satellite trajectories rarely pass on the overhead of the ground user.
	Hence, the elevation angle of the LEO satellite needs to be determined from the relative position of the satellite and the ground user in LEO S2G communications.
	For the STARLINK-4105 case, the proposed method results in good agreement with the measurements because the satellite elevation angle relative to the receiver in Manhattan was between 15$^{\circ}$ and 20$^{\circ}$. 
	For STARLINK-5240, where the elevation angle exceeded 20$^{\circ}$, a greater number of propagation paths were captured compared with the measurements. 
	Accordingly, the simulation results for the STARLINK-5240 do not fit well, but the overall trend remains similar.
	The proposed VTP can capture the impact of major scatterers, such as buildings, for different satellite orbits.
	
	A comparison of RMS delay spreads for the 6G pervasive channel model (6GPCM)~\cite{CXWang2022TVT}, the traditional method, and the proposed methods is shown in Fig.~\ref{fig:RMSDSComp}.
	The proposed method also shows good agreement with the verified 6GPCM, indicating its capability to generate accurate LEO S2G channel characteristics.
	Although the traditional method emits rays from points following the orbit, its larger angular dispersion leads to a greater delay spread compared with the proposed method. 
	While GBSMs, such as 6GPCM and QuaDRiGa, offer lower complexity by generating scatterers through statistical distributions of power, delay, and angles, they often lack site-specific information. In contrast, the computational complexity of the proposed ray-tracing method is inherently higher than that of the stochastic models because the proposed method relies on the triangulation of imported environmental data to generate exact propagation paths. Nevertheless, the proposed method is vital for LEO S2G communications, as it achieves significantly higher accuracy in capturing multipath for specific propagation environments.

\begin{figure}[t]
	\centering
	\includegraphics[width=0.9\linewidth]{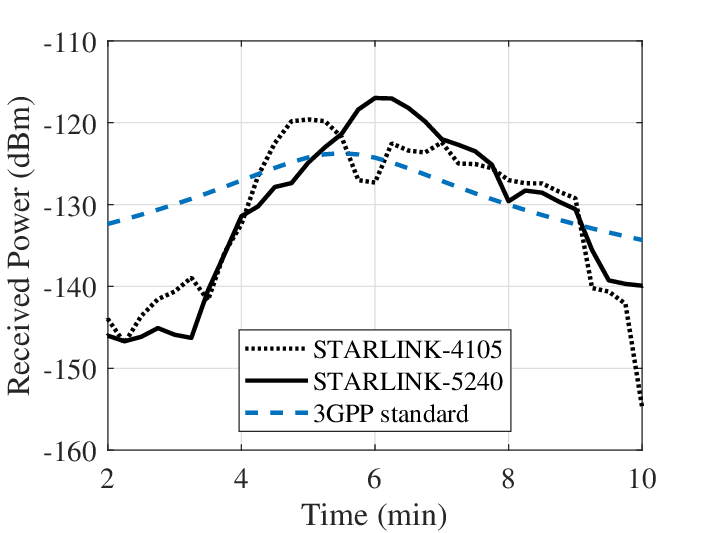}
	\caption{A comparison of the received powers among the STARLINK-4105, STARLINK-5240, and 3GPP standard \cite{3gpp2017study} in Manhattan city at 2 GHz.}
	\label{fig:3GPP}
\end{figure}

	The received powers for STARLINK-4105, STARLINK-5240, and the Third-generation partnership project (3GPP) standard \cite{3gpp2017study} in Manhattan city are shown in Fig.~\ref{fig:3GPP}.
	The transmission power of the satellite is 30 dBm, and the carrier frequency is 2 GHz.
	For the 3GPP standard, the scenario setting parameters are based on the urban scenarios because the ray-tracing simulation environment is Manhattan city.
	It is seen that the proposed method is basically consistent with the trend of 3GPP results, where the results are quite similar between the 4th and 9th minutes (the satellite elevation angle ranges from 20$^{\circ}$ to 67$^{\circ}$), where the LoS probability increases and LoS is the dominant propagation mechanism.
	Since the parameters and the path loss model of the 3GPP are fitted based on the measurements, it cannot model the blockage accurately for specific scenarios.
	Hence, there is a significant discrepancy when the elevation angle is small in Fig. \ref{fig:3GPP} because the blockage effect needs to be considered in the small elevation angle cases for LEO S2G communications.
	At low satellite elevation angles, received rays undergo up to second-order or third-order reflections, striking buildings near-vertically and creating shallow reflection angles. 
	This phenomenon drastically reduces received power at a small reflection angle and number of rays.  

\begin{figure}[t]
		\centering
		\includegraphics[width=0.8\linewidth]{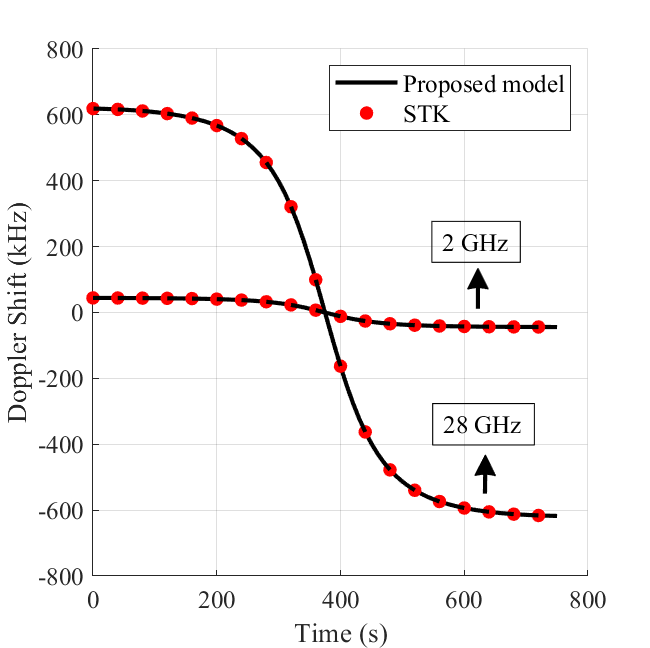}
		\caption{Doppler shifts for the proposed method and STK of the STARLINK-4105 with different time slots at 2 GHz and 28~GHz in Manhattan city.}
		\label{fig:DopplerVsTime}
	\end{figure}

\begin{figure}[t]
		\centering
		\includegraphics[width=0.8\linewidth]{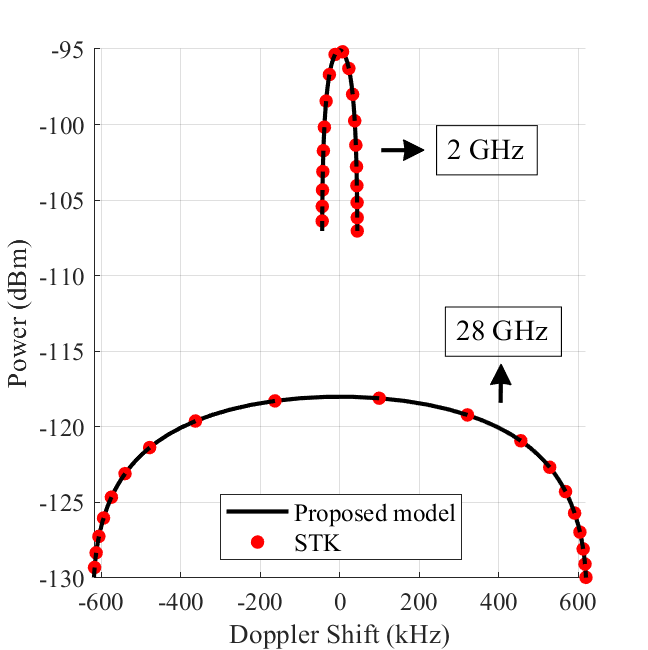}
		\caption{Doppler PSDs for the proposed method and STK of the STARLINK-4105 at 2 GHz and 28~GHz in Manhattan city.}
		\label{fig:DopplerPSD}
	\end{figure}

	The Doppler shift with operation time and Doppler PSD for the STARLINK-4105 in Manhattan city at the 2 GHz and 28 GHz are shown in Fig. \ref{fig:DopplerVsTime} and Fig. \ref{fig:DopplerPSD}, respectively.
	The results of the Doppler shift are compared with the commercial software Systems Tool Kit (STK) \cite{Abdrabou2023OJVT}, where the simulation results from the proposed method fit well with the results from the commercial software.
	For the simulation, the LoS rays always exist to keep consistency with the commercial software.
	The Doppler frequency shift decreases when the STARLINK-4105 satellite moves on top of the receiving point at 2~GHz and 28~GHz.
	At 2 GHz, the STARLINK-4105 moves to the position when the ground terminal observes a maximum elevation angle at 370 s, and the maximum Doppler shift is 44 kHz, which is close to 48~kHz in the 3GPP TR 38.811~\cite{3gpp2017study}.
	The Doppler shift is an important channel characteristic for the LEO S2G networks design.
	
\begin{figure}[t]
		\centering
		\includegraphics[width=0.9\linewidth]{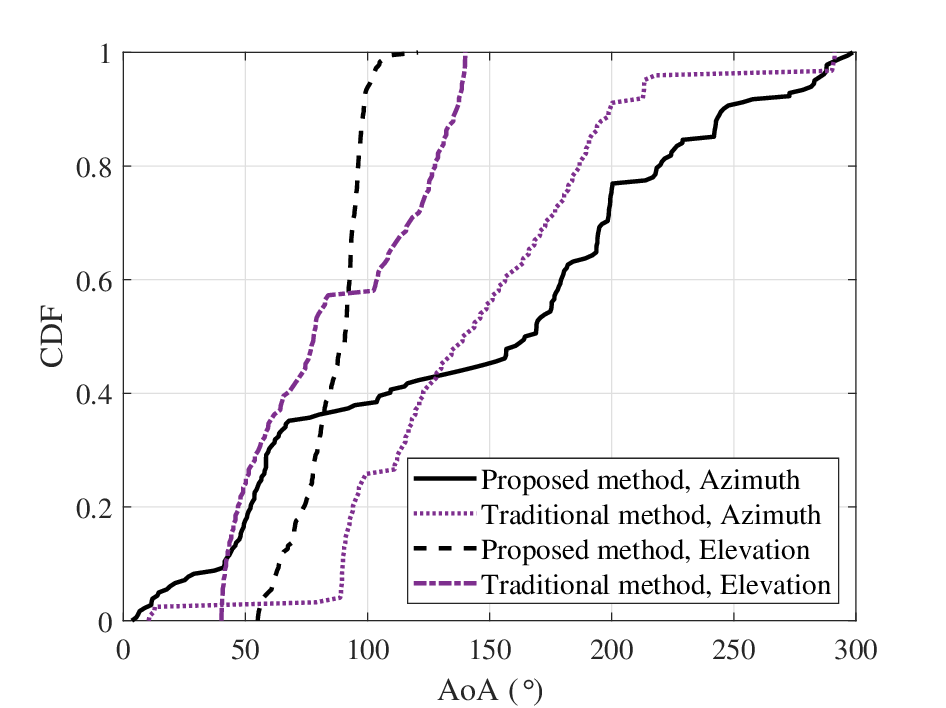}
		\caption{ AAoAs and EAoAs for the proposed and traditional methods \cite{Oestges2001TVT} of STARLINK-4105 in Manhattan city.}
		\label{fig:AoA}
	\end{figure}

	The azimuth angle of arrival (AAoA) and elevation AoA (EAoA) for the traditional and proposed methods in Case 2 are shown in Fig. \ref{fig:AoA}. 
	The AAoA results obtained from both the traditional and proposed methods show a similar range. 
	However, the distribution from the proposed method is more uniform, indicating its ability to capture more rays from each azimuth angle and better represent the channel characteristics. 
	For the EAoA, the traditional method exhibits a wider range due to greater angular dispersion in the elevation domain. 
	In contrast, the proposed method emits parallel rays from the VTP, preserving the angular characteristics of LEO S2G communications. 
	As a result, the EAoA of the proposed method aligns more closely with simulation propagation conditions, as it can be found in Fig. \ref{fig:AoA}.  
	

	\subsection{Complexity Analysis}
	We provide a comparison of computation time for the same scenario in case 2 for dynamic ray-tracing with and without the proposed pre-processing method.
	All simulations are conducted on a standard desktop equipped with an Intel Core i7-10700 2.9 GHz CPU and 32 GB RAM.
	The total computation time for the entire LEO satellite trajectory with different propagation mechanism settings is summarized in Table~\ref{table:3}, where the trajectory is discretized into 76 transmission points.
	Moreover, the reflection order is varied from 1 to 4, while the diffraction order is fixed as 1.
	From Table~\ref{table:3}, the computational time reduction of dynamic ray-tracing using the proposed pre-processing method is up to 72.57\% compared to without the proposed pre-processing method.
	The time reduction percentage of dynamic ray-tracing decreases when the reflection order increases because the number of angular ranges increases after the pre-processing stage. 
	Consequently, as the reflection order grows, the efficiency gained from ray creation and suppression diminishes, leading to an increase in the number of simulation points along the satellite trajectory.
	
\begin{figure}[t]
		\centering
		\includegraphics[width=0.9\linewidth]{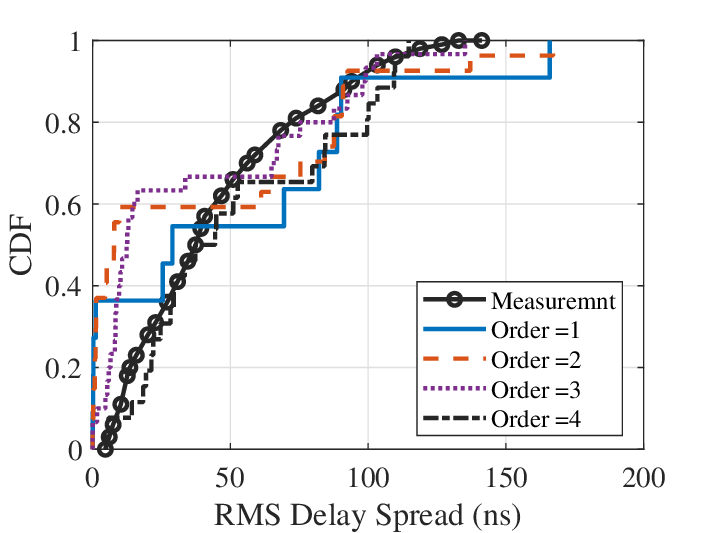}
		\caption{A comparison of RMS delay spreads for measurements \cite{Cid2016TVT} and different reflection orders in Manhattan city.}
		\label{fig:complexity}
	\end{figure}

	 In Fig. \ref{fig:complexity}, as the reflection order increases, the accuracy of the proposed channel model is also enhanced. When the reflection orders are 3 and 4, the RMS delay spreads are in good agreement with the measurements. Considering the time consumption in Table IV, the reflection order can be set to 3 for urban scenarios because it has a good trade-off between complexity and accuracy.
	
	\begin{table}[t]
		\centering
		\caption{Time consumption for different reflection-orders in test environments. }
		\label{table:3}
		\begin{tabular}{|c|c|c|c|c|}
			\hline
			Time (s)  &Order=1 &Order=2 &Order=3& Order=4 \\
			\hline
			Traditional method\cite{Oestges2001TVT}&74.58&89.44&97.42&107.31\\
			\hline
			Proposed method &20.46&36.88&45.14&60.21\\
			\hline
			\textbf{Reduction}&\textbf{72.57\%}&\textbf{58.77\%}&\textbf{53.67\%}&\textbf{43.89\%}\\
			\hline
		\end{tabular}
	\end{table}
	
\section{Conclusions}
	In this paper, a novel dynamic ray-tracing channel model for 6G LEO S2G communication systems has been proposed, addressing the critical challenges of multipath fading and long-distance propagation. 
	By integrating realistic satellite orbit dynamics through coordinate transformation and environmental data, the channel model has effectively captured the interplay of elevation angle variations and urban multipath propagation. 
	The VTP has been proposed based on the first Fresnel zone to enable accurate modeling of parallel rays near the ground in the LEO S2G ray-tracing channel model. 
	The proposed pre-processing method has significantly reduced computational complexity by combining with dynamic ray-tracing. 
	Validation through real-world measurements and simulations in roadside and urban scenarios has demonstrated the accuracy of the proposed method based on the received power, RMS delay spread, path loss, Doppler PSD, and angular characteristics. 
	To further accelerate the proposed method, the acceleration methods based on CPU and GPU need to be considered to overcome the limitations of the real-time simulation requirements and large-scale satellite constellations.

\begin{IEEEbiography}[{\includegraphics[width=1in,height=1.25in,clip,keepaspectratio] {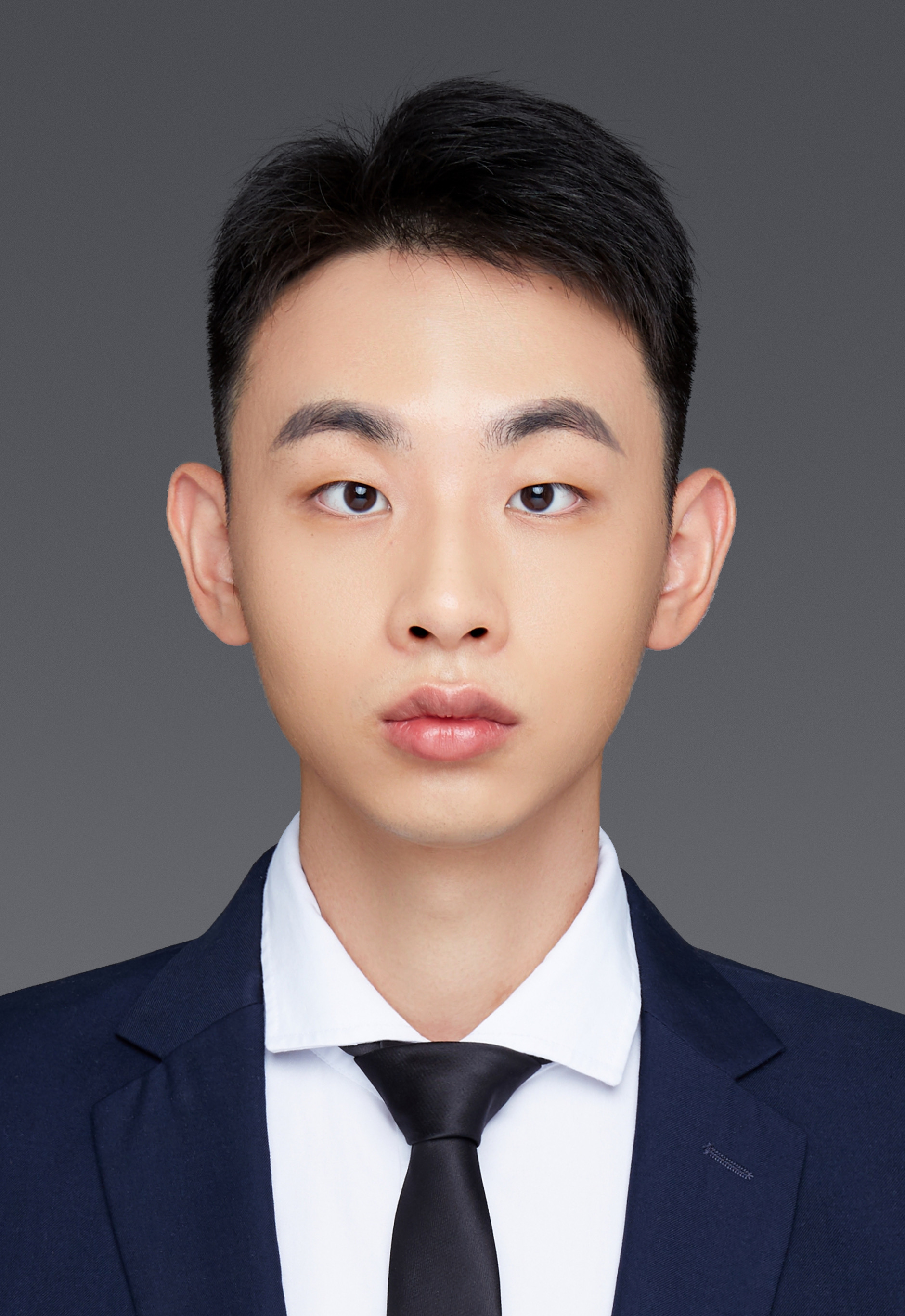}}]
	{Songjiang Yang} (Member, IEEE) is currently a postdoctoral researcher at the Pervasive Communication Research Center, Purple Mountain Laboratories, Nanjing, China. He received B.Eng. (First Class Hons.) and Ph.D. degrees in electrical and electronic engineering from the University of Sheffield, Sheffield, U.K., in 2017 and 2022 respectively. His current research interests include ray-tracing channel modeling, UAV communications, and system performance evaluation. Dr. Yang was selected for the Outstanding Postdoctoral Fellow Program in Jiangsu Province and received the Best Paper Award from WCSP 2024, IEEE ICCT 2025, and CIoTSC 2025.
\end{IEEEbiography}

\begin{IEEEbiography}[{\includegraphics[width=1in,height=1.25in,clip,keepaspectratio] {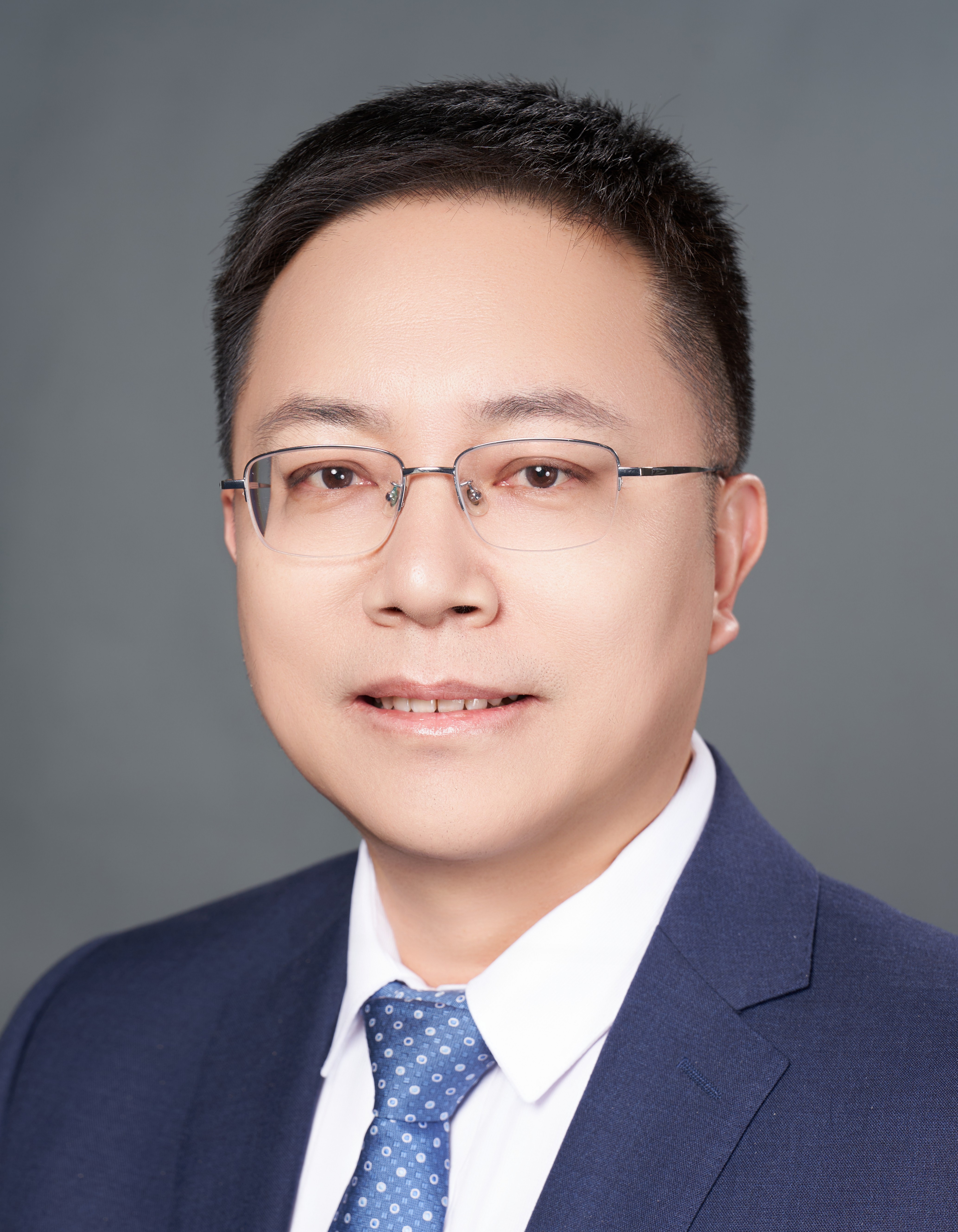}}]
	{Cheng-Xiang Wang} (Fellow, IEEE) 
	received the B.Sc. and M.Eng. degrees in communication and information systems from Shandong University, China, in 1997 and 2000, respectively, and the Ph.D. degree in wireless communications from Aalborg University, Denmark, in 2004.	
	
	He has been with Southeast University, Nanjing, China, as a Professor since 2018. He worked as the Dean of the School of Information Science and Engineering, Southeast University from 2020 to 2026, and is currently a Vice President of Southeast University. He is also a Professor with the Purple Mountain Laboratories, Nanjing. He was a Research Assistant with Hamburg University of Technology, Hamburg, Germany, from 2000 to 2001, a Visiting Researcher with Siemens AG Mobile Phones, Munich, Germany, in 2004, and a Research Fellow with the University of Agder, Grimstad, Norway, from 2001 to 2005. He was with Heriot-Watt University, Edinburgh, U.K., from 2005 to 2018, where he was promoted to a Professor in 2011. He has authored 5 books and more than 380 papers in refereed journals, including 35 highly cited papers.He has been granted over 50 invention patents, including 4 U.S. patents. He has also delivered 42 invited keynotes and 24 tutorials in international conferences. His current research interests include wireless channel measurements and modeling, 6G/B6G intelligent wireless networks, and electromagnetic information theory.
	
	Dr. Wang is a member of the Academia Europaea (The Academy of Europe) and European Academy of Sciences and Arts (EASA), a fellow of the Royal Society of Edinburgh (FRSE), IEEE, IET, and China Institute of Communications (CIC), an IEEE Communications Society Distinguished Lecturer in 2019 and 2020, and a Highly-Cited Researcher recognized by Clarivate Analytics in 2017--2020 and 2025. He was an Executive Editorial Committee Member of IEEE Transactions on Wireless Communications from 2019 to 2025. He has served as an Editor for over 16 international journals, including IEEE Transactions on Wireless Communications from 2007 to 2009, IEEE Transactions on Vehicular Technology from 2011 to 2017, and IEEE Transactions on Communications from 2015 to 2017. He was a Guest Editor of the IEEE Journal on Selected Areas in Communications, IEEE Transactions on Big Data, and IEEE Transactions on Cognitive Communications and Networking. He has served as a TPC Chair and General Chair for more than 30 international conferences. He received the IEEE Neal Shepherd Memorial Best Propagation Paper Award in 2024 and 2026. He also received 21 Best Paper Awards from international conferences.
\end{IEEEbiography}

\begin{IEEEbiography}[{\includegraphics[width=1in,height=1.25in,clip,keepaspectratio] {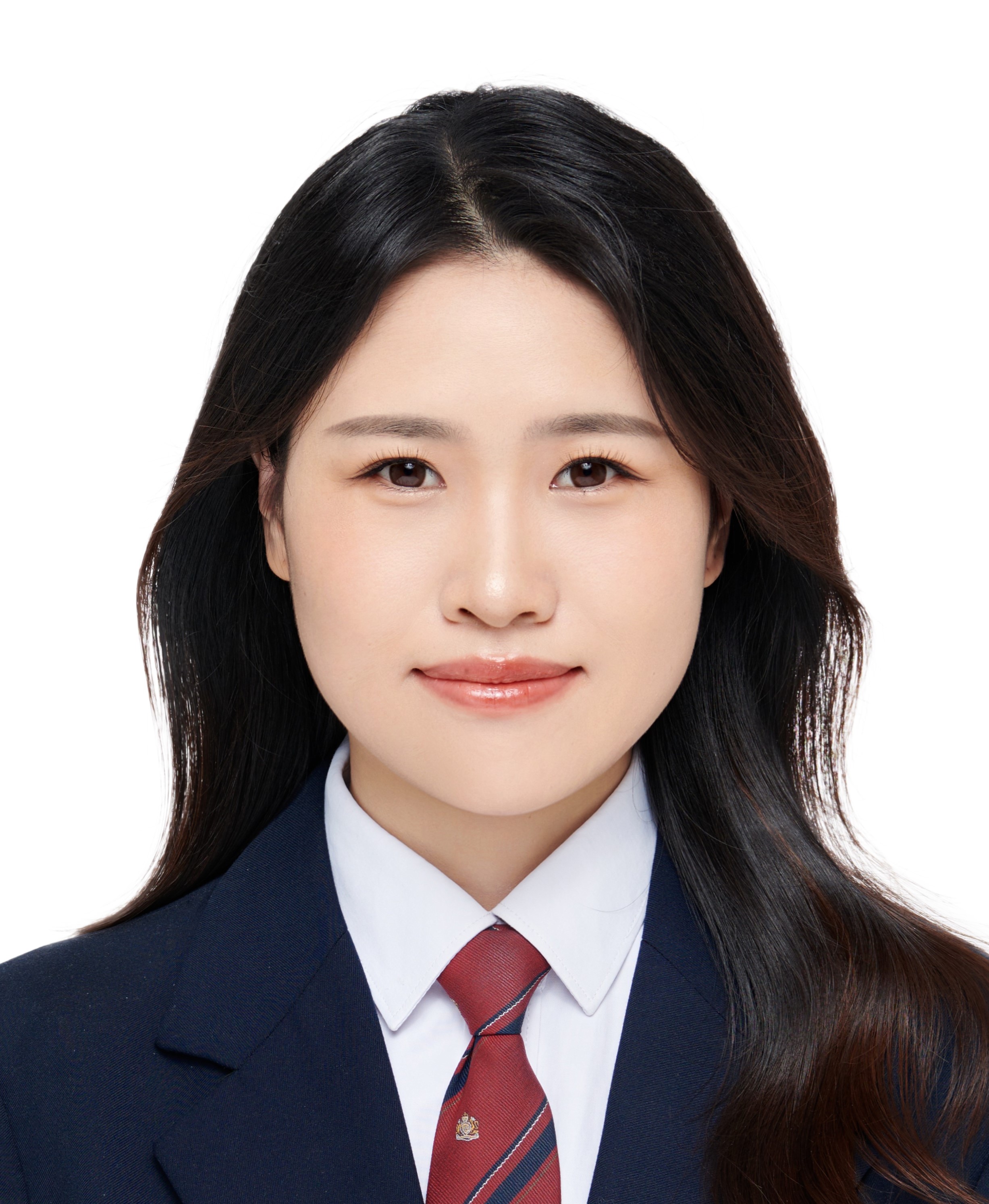}}]
	{Yinghua Wang} received the M.Sc. degree in Information and Communication Engineering from Harbin Institute of Technology, China, in 2017. She is currently a Wireless Channel Engineer in the Purple Mountain Laboratories, Nanjing, China. Her current research interests include 6G wireless channel modeling and channel simulators.
\end{IEEEbiography}

\begin{IEEEbiography}[{\includegraphics[width=1in,height=1.25in,clip,keepaspectratio] {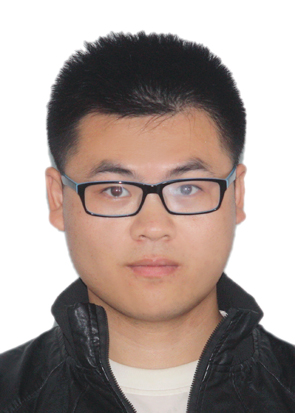}}]
	{Jie Huang} (Senior Member, IEEE) received the B.E. degree in information engineering from Xidian University, China, in 2013, and the Ph.D. degree in information and communication engineering from Shandong University, China, in 2018. From October 2018 to October 2020, he was a Post-Doctoral Research Associate with the National Mobile Communications Research Laboratory, Southeast University, China, supported by the National Postdoctoral Program for Innovative Talents. From January 2019 to February 2020, he was a Post-Doctoral Research Associate with Durham University, U.K. Since March 2019, he has been a part-time Researcher with Purple Mountain Laboratories, China. Since November 2020, he has been an Associate Professor with the National Mobile Communications Research Laboratory, Southeast University. He has authored and co-authored more than 170 papers in refereed journals and conference proceedings. His research interests include millimeter wave, massive MIMO, reconfigurable intelligent surface channel measurements and modeling, electromagnetic information theory, and 6G wireless communications. He received the IEEE Neal Shepherd Memorial Best Propagation Paper Award in 2024 and the Best Paper Awards from WPMC 2016, WCSP 2020, WCSP 2021, WCSP 2024, and IEEE ICCT 2025. He has served as an Editor for IEEE TRANSACTIONS ON GREEN COMMUNICATIONS AND NETWORKING and an Associate Editor for IEEE TRANSACTIONS ON VEHICULAR TECHNOLOGY. He has delivered more than 15 tutorials in international conferences, including IEEE Globecom and IEEE ICC.
	
\end{IEEEbiography}

\begin{IEEEbiography}[{\includegraphics[width=1in,height=1.25in,clip,keepaspectratio] {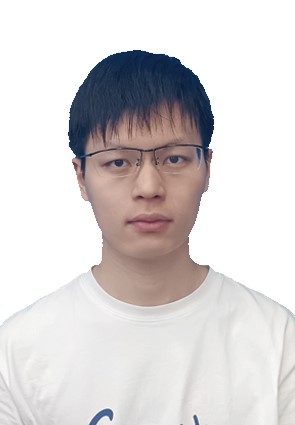}}]
	{Yuyang Zhou} received the B.Eng. degree in information science and engineering from the Southeast University, Nanjing, China, in 2022, where he is currently pursuing the Ph.D. degree. His research interests include wireless channel modeling and massive MIMO communications.
\end{IEEEbiography}	

\begin{IEEEbiography}[{\includegraphics[width=1in,height=1.25in,clip,keepaspectratio] {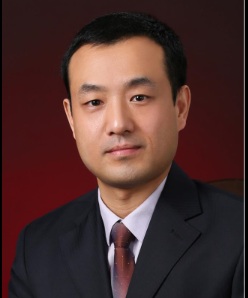}}]
	{Wei Feng} (Senior Member, IEEE)
 received the B.S. and Ph.D. degrees from the Department of Electronic Engineering, Tsinghua University, Beijing, China, in 2005 and 2010, respectively. He is currently a Professor with the Department of Electronic Engineering, Tsinghua University. He is also the Vice Dean of the Shuimu College, Tsinghua University; and the Chief Scientist of network science with the State Key Laboratory of Space Network and Communications, Beijing. His research interests include space–air–ground integrated networks, 6G mobile communications, maritime Internet of Things, and the internet of intelligent robots. He is a fellow of China Institute of Communications. He has received the National Technological Invention Award of China in 2016, the Outstanding Young Scholars Fund of Natural Science Foundation of China (NSFC) in 2019, and the Distinguished Young Scholars Fund of NSFC in 2024. He served as an Editor for IEEE TRANSACTIONS ON COGNITIVE COMMUNICATIONS AND NETWORKING from 2019 to 2023. He serves as the Assistant to the Editor-in-Chief for \textit{China Communications} and an Associate Editor for IEEE TRANSACTIONS ON AEROSPACE AND ELECTRONIC SYSTEMS.
\end{IEEEbiography}	

\begin{IEEEbiography}[{\includegraphics[width=1in,height=1.25in,clip,keepaspectratio] {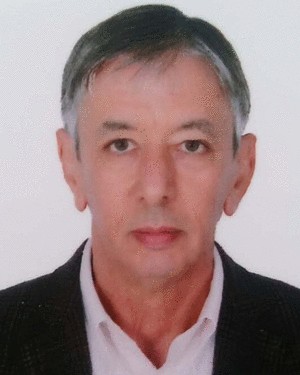}}]
	{el-Hadi M. Aggoune} (Life Senior Member, IEEE) received his Ph.D. in electrical engineering from the University of Washington (UW) in Seattle, WA, USA. He has served at several universities in the US and abroad across various academic ranks, including as an Endowed Chair Professor. He is listed as an inventor on several patents assigned to universities and corporations, including The Boeing Company. A registered Professional Engineer in the state of Washington, he has co-authored papers in IEEE and other prominent journals and conferences, while also serving on numerous editorial boards and technical committees. Dr. Aggoune was the recipient of the IEEE Professor of the Year Award at UW and previously directed a laboratory that received the Boeing Supplier Excellence Award. He currently serves as a Professor and Director of the AI and Sensing Technologies Research Center at the University of Tabuk. His research interests include wireless communication, power systems, and artificial intelligence.
\end{IEEEbiography}	

\end{document}